\documentclass[draft]{agujournal2019}
\usepackage{url} 
\usepackage{lineno}
\usepackage[inline]{trackchanges} 
\usepackage{soul}
\draftfalse

\begin{document}

%
%

\title{The role of freshwater forcing on surface predictability in the Gulf of Mexico}

%
%




\authors{Daoxun Sun\affil{1}, Annalisa Bracco\affil{1} and Guangpeng Liu\affil{1}}


\affiliation{1}{School of Earth and Atmospheric Sciences, Georgia Institute of Technology}




\correspondingauthor{Daoxun Sun}{sdxmonkey@126.com}




\begin{keypoints}
\item LCS and MARS show two-way interaction related with river representation and resolution.
\item Spread of SST and SSS show strong, but opposite, seasonalities.
\item Increasing resolution decreases the predictability of surface relative vorticity.
\end{keypoints}

%
%

%
%


\begin{abstract}

The predictability of fields at the ocean surface in the northern Gulf of Mexico (GoM) is investigated through five ensembles of regional ocean simulations between 2014 and 2016. The ensembles explore two horizontal resolutions and different representations of the riverine inflow, and focus on the Loop Current system (LCS) and the Mississippi-Atchafalaya River System (MARS) interactions.

The predictability of the surface fields is high in the northern GoM if the atmospheric forcing and the flow at Yucatan Channel are known, and the ensembles simulate similar LCS behavior up to 5 months. In terms of LCS-MARS interactions, the ensembles confirm that they are strongly modulated by the LC mesoscale variability. The relationship is two-ways, with the LCS being influenced by - and not only influencing - the freshwater plume. Whenever the freshwater flux is strong, the northward extension of the LCS is constrained. The ensemble simulations also indicate that this influence is stronger if the riverine inflow is simulated in an active fashion with a meridional velocity component proportional to the flux. Sea surface temperature (SST) and salinity (SSS) predictability have opposite seasonality in their signal, with the SST (SSS) field being more predictable in summer (winter). Partially resolving submesoscale instabilities and improving the accuracy of the riverine fluxes' representation causes the spread to increase, especially in SST. Finally, the predictability of surface relative vorticity decreases in amplitude when increasing resolution due to feedbacks between the mesoscale and submesoscale circulations, but retains most of its intraseasonal and interannual signal.

\end{abstract}

\section*{Plain Language Summary}
In this work we ask the questions: how predictable is the circulation and the sea surface fields in the northern Gulf of Mexico (GoM), and what is the role of the riverine input contributed by the the Mississippi-Achafalaya River System? Under the assumption of a perfect model and known atmospheric forcing and oceanic boundary fields, we simulate the GoM circulation through five ensembles which members differ only in their initial conditions. The ensembles span 32 months and explore two horizontal resolutions and different representations of the riverine inflow. 

It is found that the predictability of the main current in the GoM, the so-called Loop Current, is high under the conditions of our simulations. The northward extension of the Loop Current, however, is influenced by the riverine input and its representation. This influence is stronger if the riverine inflow is simulated accounting for a meridional velocity component proportional to the freshwater flux. The sea surface temperature (SST) field is more predictable in summer, and the salinity (SSS) in winter. Increasing model resolution, on the other hand, decreases the predictability due to feedbacks between the mesoscale (several tens of kilometers in scale) and submesoscale (less than five kilometers in scale) circulations.

\section{Introduction}
\label{sec:Introduction}
The northern Gulf of Mexico (GoM) is a multifaceted system whose spatial and temporal variability is modulated by large mesoscale circulations, strong currents, energetic submesoscale dynamics and abundant river discharge  in spring and summer \cite{cardona2016predictability,luo2016submesoscale,barkan2017submesoscale1}.

The prevalent mesoscale pattern is the Loop Current (LC) that enters the Gulf between Cuba and the Yucatan peninsula and contributes Atlantic waters with relatively high temperatures, low salinity and low nutrients.
At irregular intervals, ranging from 0.5 to 18 months \cite{leben2005altimeter}, the LC sheds large anticyclonic mesoscale eddies (200 – 300 km in diameter), also known as Rings,  often surrounded by smaller mesoscale vortices, both cyclonic and anticyclonic, and intense vorticity filaments \cite{sturges2000frequency}. 
The LC eddy shedding is a  nonlinear process during which the LC usually changes from a northward extending position that approaches the continental slope of the northern GoM to a southward retracted position in the Yucatan Channel \cite<e.g.,>[]{oey2005loop}. 
The LC eddies then translate westward with a velocity of about ~2–5 km d$^{-1}$ and have lifetimes from few months to about a year \cite<e.g.,>[]{vukovich2007climatology}. 

The LC system or LCS includes the LC, the Rings and the smaller mesoscale eddies, cyclonic and anticyclonic, surrounding the LC and the Rings. It influences the northern GoM circulations and processes on the continental slope and  shelfbreak  \cite{schiller2011dynamics,cardona2016predictability}, causing persistent, strong, cross-isobath transport of tracers and materials, including river discharge \cite{hamilton2005eddies,hu2005mississippi,schiller2014loop}. Riverine waters enter the GoM from several rivers - more than 17 - with the Mississippi and its Atchafalaya diversion being the main contributors, often grouped together as Mississippi-Atchafalaya River system (MARS). 

The Mississippi-Atchafalaya is the largest river in North America with an average rate surpassing 14,000 $m^3 s^{-1}$ \cite{hu2005mississippi}. It represents 80\% of the annual freshwater input to the Gulf of Mexico, 90\% of the total nitrogen load (mainly of agricultural origin) and 87\% of the phosphorous load to the basin \cite{dunn1996trends}. 
\citeA{schiller2011dynamics} have shown that offshore transport associated with the interaction of the river plume and the mesoscale circulations happens frequently and in all seasons, and depends on the mesoscale variability and on topographic and wind effects.
Furthermore, the juxtaposition of mesoscale eddies and filaments in waters with very different densities contributes to frontal and baroclinic instabilities at scales of few kilometers and to the formation of fronts and lateral convergence zones \cite{zhong2012pattern,zhong2013submesoscale,luo2016submesoscale,barkan2017submesoscale1,barkan2017submesoscale2,d2018ocean,androulidakis2018influence}. 
Submesoscale circulations influence mixing processes within and across the mixed layer \cite{liu2021submesoscale}, and their impacts vary  first and foremost seasonally, as shown through various modeling efforts \cite<e.g.,>[]{zhong2012pattern,luo2016submesoscale,bracco2019mesoscale,liu2021submesoscale} and drifter deployments  \cite{poje2014submesoscale,d2018ocean}, but also diurnally and interannually \cite{sun2020diurnal}.

Understanding what controls the penetration of the LC and the Rings into the GoM and the representation in ocean models is  key to better predictions, and has invaluable societal benefit. Constraining the predictability potential of the LC matters for any kind of fishery management, for the oil and gas industry operation and disaster preparedness, for improving forecasts of hurricanes and tropical storms, and for better constraining  heat and moisture fluxes from the GoM into the continental U.S. 

In this work we focus on the predictability potential of the LCS in its linkages with the MARS in the northern GoM. 
We analyze the predictability potential of surface fields, and particularly sea surface height, temperature and salinity, and relative vorticity. 
Our goal is to identify the processes that matter for their variability and modes of interaction, and their representation in a regional ocean model across a period of nearly three years, answering the question: if, how and how much is the LC system influenced by the freshwater plume in submesoscale permitting and mesoscale resolving simulations? 
We do so using ensembles of simulations with the same model and exploring the ensemble spread under different  representations of the riverine input and model resolutions, considering mesoscale resolving and submesoscale permitting ensembles. In our evaluation, the ensemble spread provides an upper bound to the predictability potential.

Background information on the period considered and a brief summary of relevant literature on the LC-MARS dynamics is presented in Section \ref{sec:Background}.
Details on the modeling set up and a short synthesis of the climatology of the ensemble simulations are provided in Section \ref{sec:Model}. 
The potential predictability is quantified in Section \ref{sec:Results}, with a discussion focused on the LC evolution among different ensembles (\ref{sec:Results:LC}) and on the surface fields  (\ref{sec:Results:spread}).
Section \ref{sec:Discussion} concludes the paper with a discussion of our findings.

\section{The LC-MARS dynamics in 2014-2016}
\label{sec:Background}

\begin{figure}
\includegraphics[width=1\linewidth]{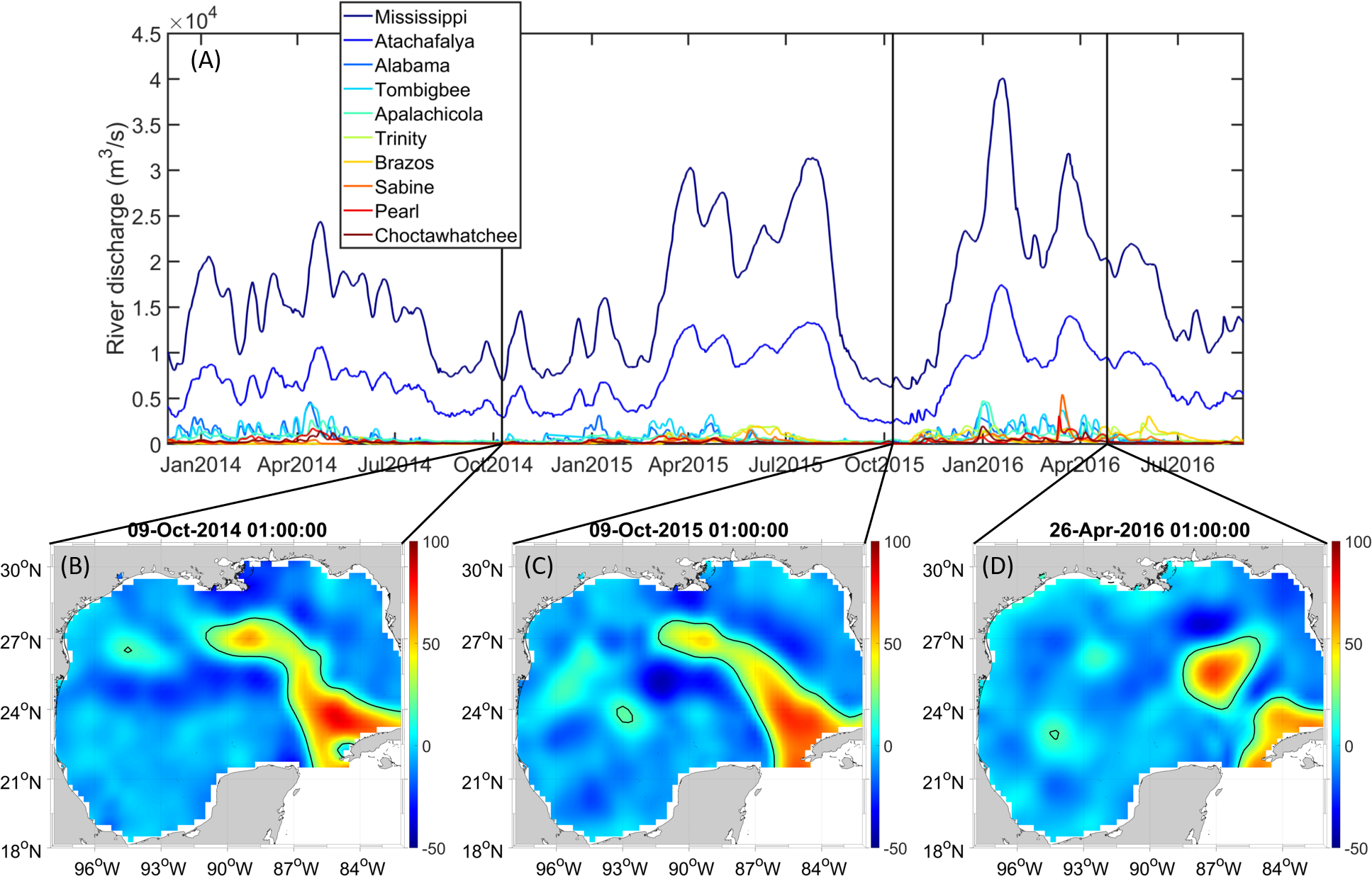}
\caption{(A): Time-series of river discharge for the 10 largest rivers of the MARS. (B-D): Satellite based sea surface height anomalies (SSHa) from the Gulf of Mexico Coastal Ocean Observing System (GCOOS) \cite{leben2002operational} on October 9th 2014 (B), October 9th 2015 (C) and April 26th 2016 (D). The 17 cm sea surface height anomaly (SSHa) contour line is highlighted in black.}
\label{fig:river_SSH}
\end{figure}

In the GoM, the interaction between the river system and the LCS is an interesting case of cross-marginal transport of riverine waters, where the physical and biological connectivity between remote offshore and shelf regions is modulated by mesoscale and submesoscale circulations spanning a wide range of scales, from hundreds of kilometers to few hundreds of meters \cite{schiller2011dynamics,schiller2014loop,barkan2017submesoscale1}. A careful analysis of these interactions using a high resolution (2 km horizontal resolution) data assimilative model, in-situ observations and satellite measurements is presented in  \citeA{androulidakis2019offshore} limited to the 2015 summer. 

The LC can be often found in a retracted state in the southern portion of the Gulf, or in an extended one reaching the Louisiana–Mississippi shelfbreak before looping and returning to Florida. 
In the extended configuration the interactions with the MARS plume are substantive, as occurred in June and July 2015. 
As mentioned, the LC transit is also accompanied by the formation of large anticyclonic eddies, the Rings, and by smaller mesoscale eddies of both sign, cyclonic and anticyclonic, through instabilities and interactions with the bathymetry \cite{hamilton2009topographic}. 
LC eddies, whenever formed, tend to travel through the Sigsbee Deep with a vertical extent of 800-1000 m \cite{hamilton1990deep,lee2003numerical}, and may interact with the surface flow in the north west of the basin while dispersing anticyclonic vorticity \cite{dimarco2005statistical} and advecting MARS plume waters at their periphery \cite{androulidakis2019offshore}.
Indeed,  \citeA{schiller2014loop}  showed with a numerical investigation that in the region north of 28$^oN$ the freshwater transport is controlled almost entirely by the LC intrusion whenever it approaches the Mississippi River mouth, by a combination of mesoscale circulations and winds if the LC or the Rings are further off-shore, and is not significantly impacted by the LC system if this is confined to the south and away from the Mississippi delta. 

In relation to submesoscale dynamics, the role of the riverine discharge is, on one hand, to enhance submesoscale currents by providing lateral buoyancy gradients, and, on the other, to suppresses them by increasing stratification near the surface.
In the GoM, suppression is predominant in winter when baroclinic instabilities are the most abundant both around and inside the LC and the Rings \cite{mensa2013seasonality,callies2016role}, but their strength and number is somewhat constrained by the freshwater input, due to the increased stratification near the surface \cite{barkan2017submesoscale1}. 
By contrast, in late spring and early summer, when the MARS is commonly at its maximum load, submesoscale enhancement by river discharge prevails because frontal dynamics are amplified by the density gradients fueled by the water of riverine origin, despite the shallow mixed layer. Observations during the Deepwater Horizon disaster and at the Taylor Energy site have underlined the importance of these riverine-induced fronts as transport pathways \cite{le2012surface,androulidakis2018influence}, while numerical simulations have shown that if the river discharge is small or null, the formation of submesoscale circulations is substantially reduced in summer and slightly enhanced in winter \cite{luo2016submesoscale}. 

We consider the period between December 2013 and August 2016, covering a year of relative low discharge (2014), a year of ‘normal-to-high’ discharge with a relative maximum in late spring as commonly observed (2015) and a year characterized by extreme flooding conditions in January-February due to the influence of the El Ni\~no Southern Oscillation (2016) (Fig. \ref{fig:river_SSH}).
Several LC intrusion episodes occurred in this timeframe. The LC extended northward towards the Mississippi river mouth and the Florida Panhandle in fall 2014 into the winter of 2015 and again in fall 2015, while a long-lived Ring detached in spring 2016, moving westwards slowly and retaining its identity through most of the year. 

In terms of predictability studies relevant to our approach, most research so far concentrated on the LC system with the objective of predicting LC eddy shedding. 
Remote sensing products and a variety of ocean models all integrating the primitive equations have been used to evaluate the predictability of the LC location, extension and overall shedding behavior from the seminal works by \citeA{oey2005wetting} and \citeA{yin2007bred} using the Princeton Ocean Model (POM), to those by \citeA{counillon2009high} using the Hybrid Coordinate Ocean Model (HYCOM), \citeA{mooers2012final} within the Gulf of Mexico 3-D Operational Ocean Forecast System Pilot Prediction Project (GOMEX-PPP), and \citeA{gopalakrishnan2013state} with the Massachusetts Institute of Technology General Circulation Model (MITgcm). 
All these studies focus on few specific shedding events and lack in generality.  

Here we are interested on the predictability potential of the LC extension in relation to its interactions with the MARS, which has been mostly unexplored, and we use ensembles performed with a regional primitive equations ocean model to determine the role of the riverine input and its modeling representation in conjunction with submesoscale variability in the absence of any data assimilation. The spread of our ensembles provides an upper bound to the predictability potential. If the model was perfect and perfectly initialized, and the forcing fields were without error or uncertainty,  the ensemble spreads would quantify the internal variability intrinsic to the GoM basin given the slightly different initial conditions used (see below).  We note that if ensemble forecasting has been recently shown to expands the prediction horizon for ocean mesoscale \cite{thoppil2021ensemble}, increasing model resolution to capture submesoscale circulations may have the opposite effect \cite{sandery2017ocean}.

\section{Model set-up and simulation details}
\label{sec:Model}
\begin{figure}
\includegraphics[width=1\linewidth]{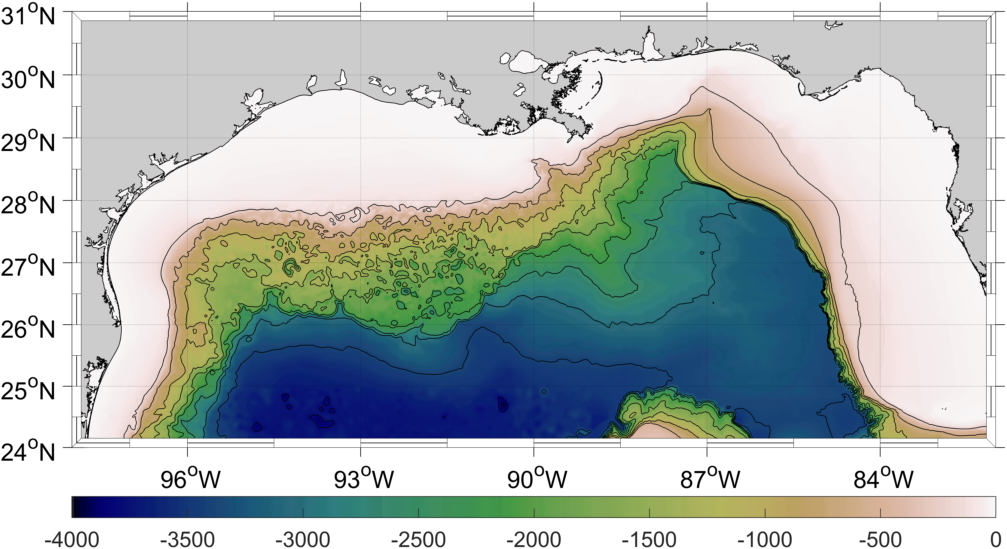}
\caption{Model domain of the northern Gulf of Mexico. Shading indicates the depth of model grids. Contour lines are plotted every 400 m.  Simulations are preformed at 1 km and 3.5 km horizontal resolution.}
\label{fig:domain}
\end{figure}

We use CROCO (Coastal and Regional Ocean COmmunity model), a regional oceanic modeling system built upon ROMS (Regional Ocean Modelling System) in its Agrif version \cite{auclair2018some}. Our model domain cover the  Gulf of Mexico north of 24\textdegree N (Fig. \ref{fig:domain}).
Boundary conditions are imposed at the southern and eastern boundaries.
A third-order upstream biased advection scheme is used for horizontal advection \cite{shchepetkin2005regional}. 
Laplacian horizontal mixing is activated for momentum, and bi-laplacian horizontal mixing is activated for tracers, while the K-profile parameterization \cite<KPP,>[]{large1994oceanic} is used for vertical mixing.
Ten harmonic constituents (M2, S2, N2, K2, K1, O1, P1, Q1, Mf, Mm) of tides from TPXO-7 global tidal model \cite{egbert2002efficient} are included.

\begin{figure}
\includegraphics[width=1\linewidth]{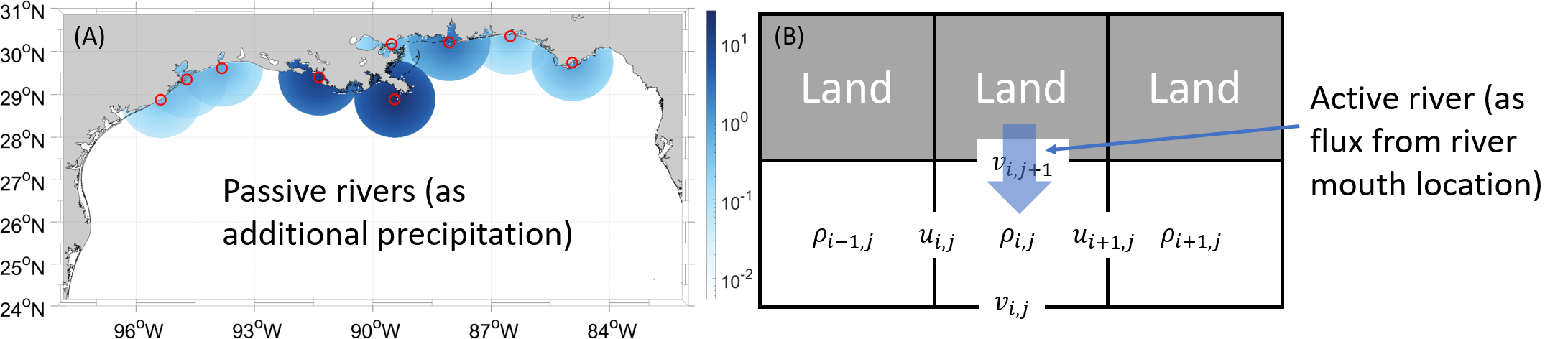}
\caption{Schematic diagram of the river configurations used in the simulations. In the passive river configuration (A), the total discharge is applied as an additional precipitation flux over the ocean grid points near the river mouth location as indicated. In the figure the flux values used are the average of April, 2015. The red circles indicate the river mouth locations. In the active river configuration (B), river discharge is imposed as a southward volume flux from the north side of the ocean grids immediately adjacent to the river mouth locations.}
\label{fig:river_scheme}
\end{figure}

Initial and boundary conditions are from the HYCOM - NCODA (Hybrid Coordinate Ocean Model - Navy Coupled Ocean Data Assimilation) analysis at 1/25\textdegree resolution \cite{cummings2005operational,cummings2013variational} (exp 31.0 for periods before April 4th, 2014, and exp 32.5 for April 4th, 2014 onward).  
CROCO is forced by 10 m wind, surface shortwave radiation, surface downward long wave radiation, precipitation, 2-m air temperature and relative humidity from the NAVy Global Environmental Model (NAVGEM), which are also used in HYCOM-NCODA runs, and the bulk parameterization is activated for the surface air-sea fluxes parameterization, again following the HYCOM-NCODA set-up.
River discharges are imposed based on the U.S. Geological Service (USGS) and the U.S. Army Corps of Engineers (USACE) daily data collected as part of the Gulf of Mexico Coastal Ocean Observing System (GCOOS) project. 
The ten rivers with the largest discharge of MARS are included in the simulations. 
In HYCOM - NCODA, the rivers are treated as a bogus surface precipitation affecting the upper 6 m of the ocean, and a three-Dimensional Variational (3DVAR) assimilation system is applied to assimilate available satellite and in situ observations.
In our simulations, the riverine forcing is implemented in an active or passive way (Fig. \ref{fig:river_scheme}). 
In the passive river configuration, the total discharge is applied as a bogus precipitation flux over several ocean grids near the river mouth location, as done in recent process studies \cite{barkan2017submesoscale1,liu2021submesoscale}. 
The passive implementation ensures that the freshwater reaches realistically far from the coast but neglects the momentum flux associated with the river outflow.
In the active river configuration, river discharge is imposed as a southward volume flux from the north side of the ocean grids adjacent to the river mouth locations. 
The salinity of the river contribution is held constant at 4 psu, and the temperature is set at the same value of the air temperature at the river mouth. The freshwater flux is largest at the surface and decays exponentially away from it towards the ocean bottom. 
Southward momentum associated with the fresh water flux is added to the v-component of the current at different vertical levels.

All simulations are performed with 70 sigma layers, to resolve the first and second baroclinic modes \cite{stewart2017vertical}. The ensembles performed cover 8 periods from December 2013 to August 2016. In each period, the circulation of the northern GoM is simulated for 5 months, focusing on different seasons, conditions of the LC system and MARS discharge flow rates, starting at the beginning of December  2013, 2015, 2016, April  2014, 2015, 2016 and August 2014 and 2015. 
Each ensemble explores both river treatment and horizontal resolution dependency, and includes 8 runs at 1 km horizontal resolution (SP for submesoscale permitting), 3 of which with the MARS runoff in the active configuration, 3 in the passive configuration and 2 runs without any river runoff, and 6 runs performed at 3.5 km horizontal resolution (MR for mesoscale resolving), 3 of which with the MARS runoff in active configuration and 3 in the passive configuration. 
Ensemble members in the same configuration group differ in their initial conditions chosen to be 1 day apart (i.e. December 1st, 2nd and 3rd). The initial spread is computed as difference among the three days, and day 1 of the ensemble simulations corresponds to the first with common forcing and boundary conditions (i.e. December 4th).

Ensemble spread has been widely used in climate and oceanography studies to estimate the potential predictability of a forecast system. Under a certain forecast range, a low (high) ensemble spread is usually accompanied by a high (low) forecast skill \cite<e.g.,>[]{murphy1988impact,barker1991relationship,buizza1997potential,moore1999dynamics}.
In the following, within each configuration group, the standard deviation (STD) among ensemble members differing only in their initial conditions has been calculated at each grid point as a way to quantify the spreading, thus inferring predictability. Time-series are then constructed by averaging the STD over the model domain. The location of LC and the Rings is identified using sea surface height anomalies (SSHa) and specifically the 17-cm SSH contour. 
This metric has been widely used in previous literature, and the contour closely matches the edge of the high-velocity core \cite<e.g.>[]{leben1993tracking,donohue2016loop,hamilton2016loop}.
The  basin average SSH is removed before detecting the 17-cm contours, to remove the basin-scale tidal signal.
We only consider contours with perimeters longer than 300 km to focus on the main LC or the large Rings.
If multiple contours are captured (e.g. the LC and a Ring at the same time), the contours extending furthest to the north is used to identify the northward extension of the LCS.

\begin{figure}
\includegraphics[width=1\linewidth]{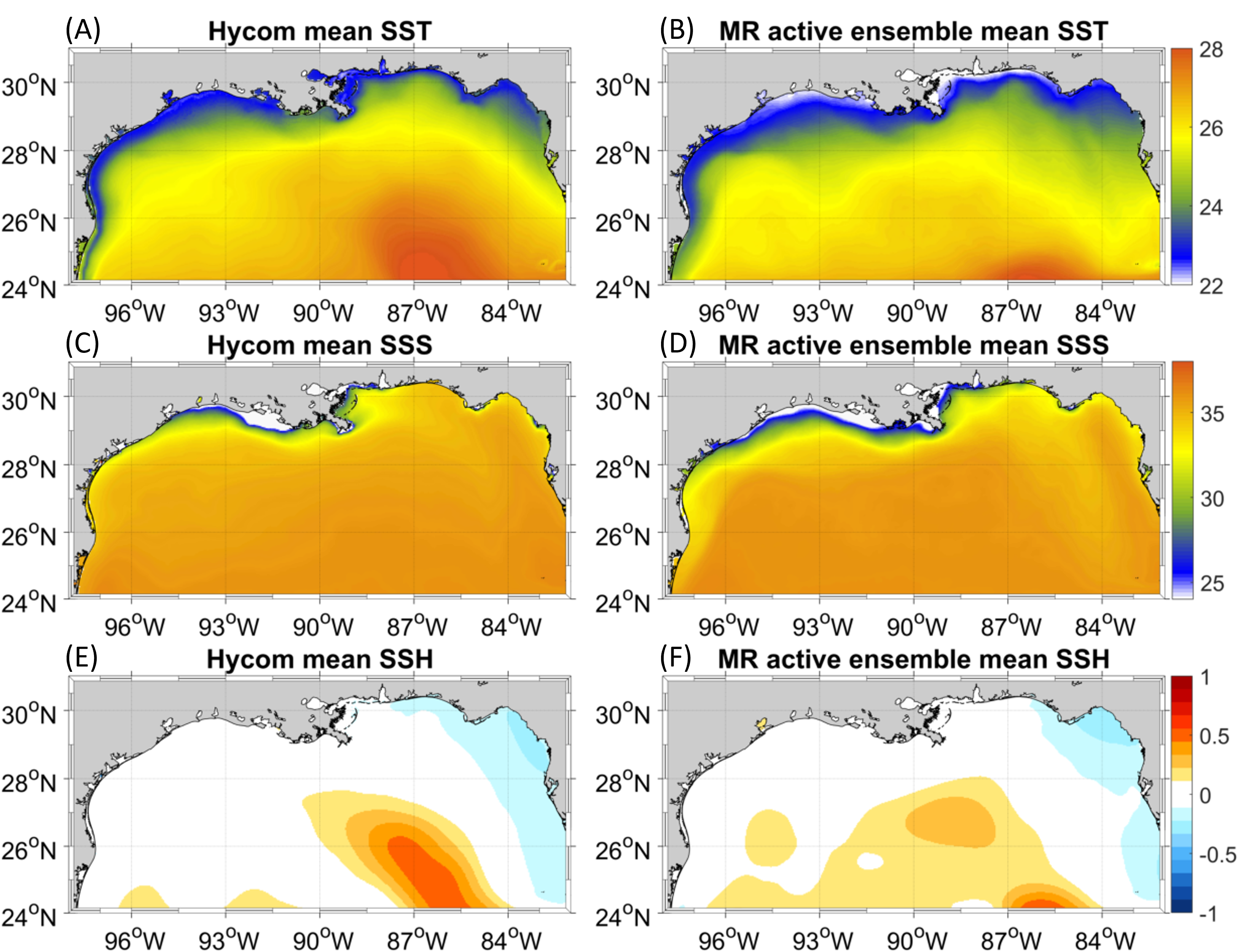}
\caption{Mean sea surface temperature (SST, A-B), sea surface salinity (SSS, C-D) and sea surface height (SSH, E-F) over the period 12/02/2013  - 08/10/2016 in HYCOM (A-C-E) and one member from the 3.5 km active configuration  ensemble (B-D-F).}
\label{fig:surf_clim}
\end{figure}

\begin{figure}
\includegraphics[width=1\linewidth]{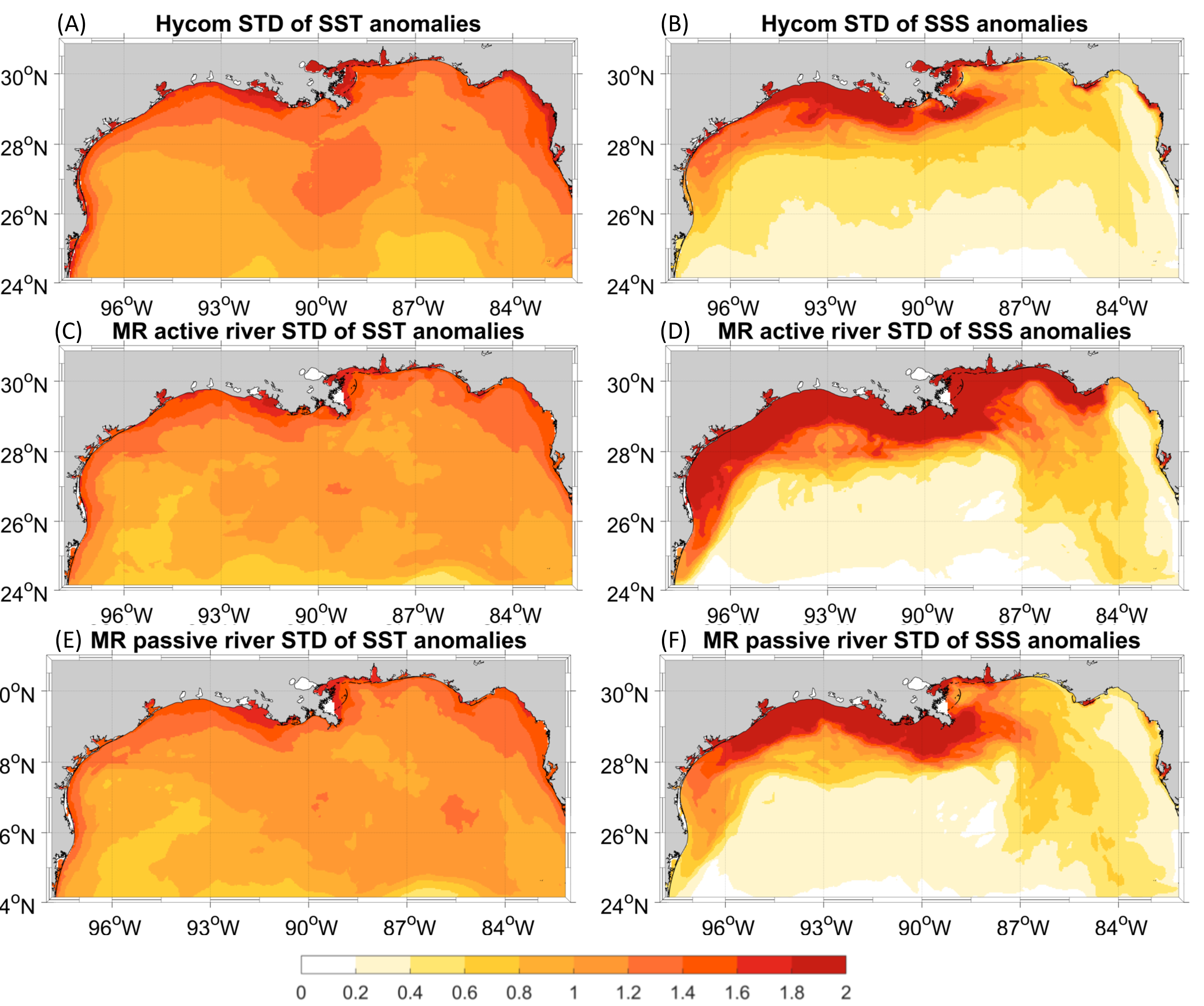}
\caption{STD of SST anomalies (A-C-E) and SSS anomalies (B-D-F)  over the period 12/02/2013  - 08/10/2016 in HYCOM (A-B), one the 3.5 km simulations with a passive river configuration (C-D) and one member from the MR active river configuration ensemble (E-F). The anomalies are calculated by removing the monthly mean climatology.}
\label{fig:surf_STD_clim}
\end{figure}

Fig. \ref{fig:surf_clim} compares mean surface fields between HYCOM data and the ensemble simulations (one member from MR active river configuration is shown here as an example. We chose to show the MR case because its resolution is close to that of HYCOM).
The simulations reasonably represent the spatial structure of the HYCOM analysis. For a brief validation of the representation of  stratification in CROCO, the reader is referred to \cite{liu2021submesoscale}.
Warm and salty water enters the basin from the southern boundary with the LC, while relatively cold and fresh water can be found along the coast.
However, sea surface temperatures (SSTs) in the simulations tend to be colder in the north and east comparing to HYCOM data, while the sea surface salinity (SSS) field is saltier in the west and fresher in the east.
In CROCO the large Rings tends to detach from the LC earlier than observed and move towards the west, while the LC remains in its extended extended configuration for longer in the HYCOM analysis. We note that the length of periods spent by the LC in an extended configuration in fall 2014 into spring 2015 was underestimated by all forecasts as well \cite{NAS2018report} and may be an intrinsic bias of ocean models \cite{liu2021GRL, Sheremet2021JPO}. 
CROCO simulations capture also the observed standard deviation (STD) patterns of the surface fields (Fig. \ref{fig:surf_STD_clim}).
The largest SST variations are found in the northern portion of the basin, where the water column is no more than 50 m deep, and in correspondence of the (mean) edge of LC shown in Fig. \ref{fig:surf_clim} E-F, where small differences in the representation of the largest mesoscale pattern in the basin causes major differences in surface temperature, especially in fall and winter.
The largest SSS variations, on the other hand, are found along the coast where the riverine freshwater enters the basin, and are greatest in the runs with an active river configuration. In CROCO the advection of salinity anomalies introduced by the Mississippi River along the LC edge is reflected in the west-east gradient of the STD pattern, and this pattern is accentuated, in both runs, compared to HYCOM.

\section{Results}
\label{sec:Results}
\subsection{The Loop Current  in the  five ensembles}
\label{sec:Results:LC}
A first key element to the predictability of the northern GoM relates to the position and evolution of the LC and the Rings. In our ensembles, the model domain extends only to 24${^o}$N, and the oceanic forcing associated to the flow at Yucatan Strait is imposed identically to all simulations as boundary condition (for a discussion on the role of Yucatan Strait forcing see e.g. \cite{ezer2003variability}. Each ensemble but for that without riverine inflow (SP no-river case) includes three members with identical resolution and river configuration, and we consider the 8 simulation periods. In each period, the initial ensemble spread is identical among ensembles (with the exception of the  no-river group) and given by the difference among the three consecutive days  in HYCOM-NCODA used as initial conditions. 

\begin{figure}
\includegraphics[width=1\linewidth]{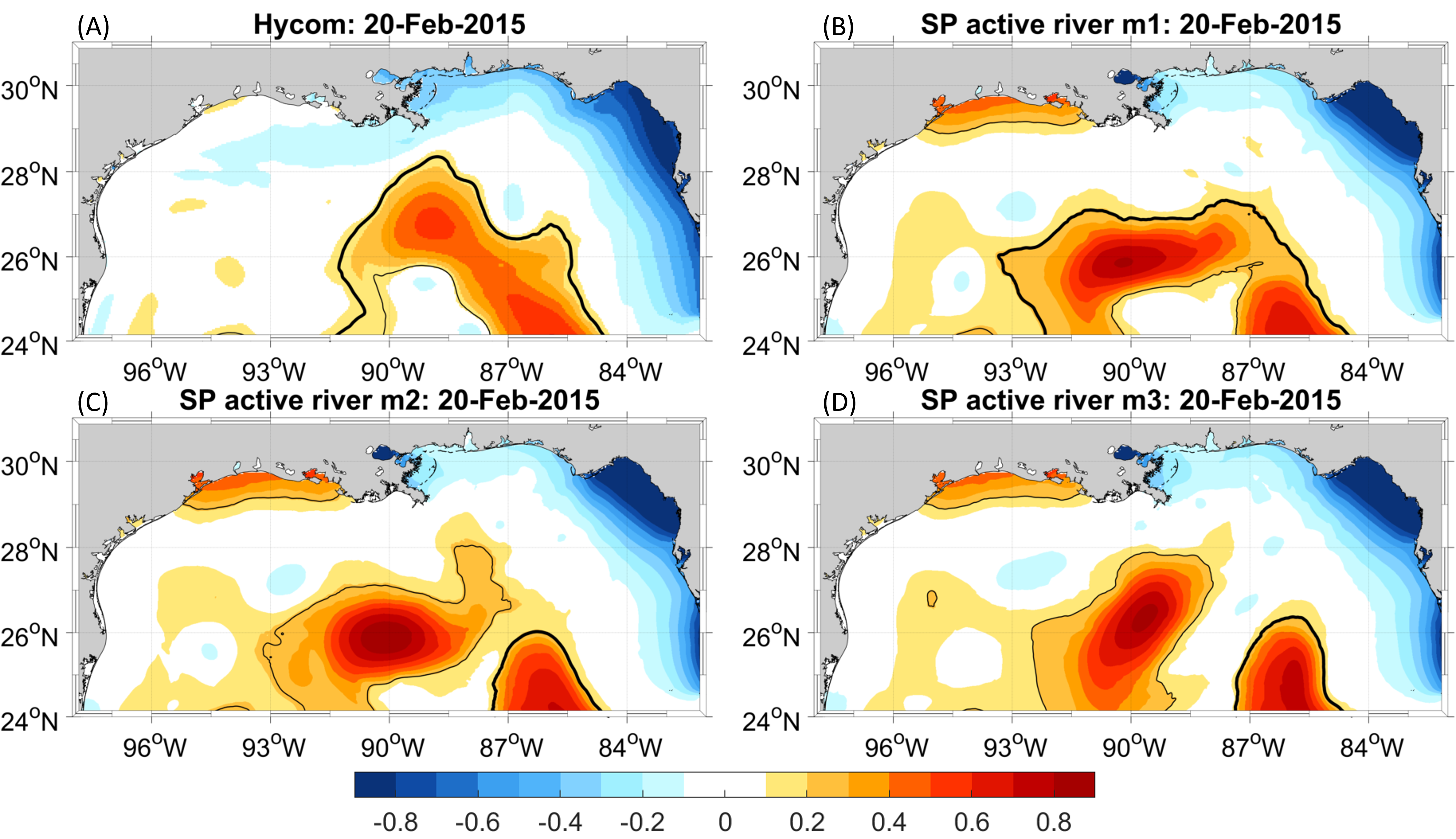}
\caption{SSH in Hycom (A) and in the 3 members of SP active river runs (BCD) on Feb. 20, 2015. The black lines follow the 17 cm SSHa contours and the thick line identifies the LCS location.}
\label{fig:SSH_LC}
\end{figure}

\begin{figure}
\includegraphics[width=1\linewidth]{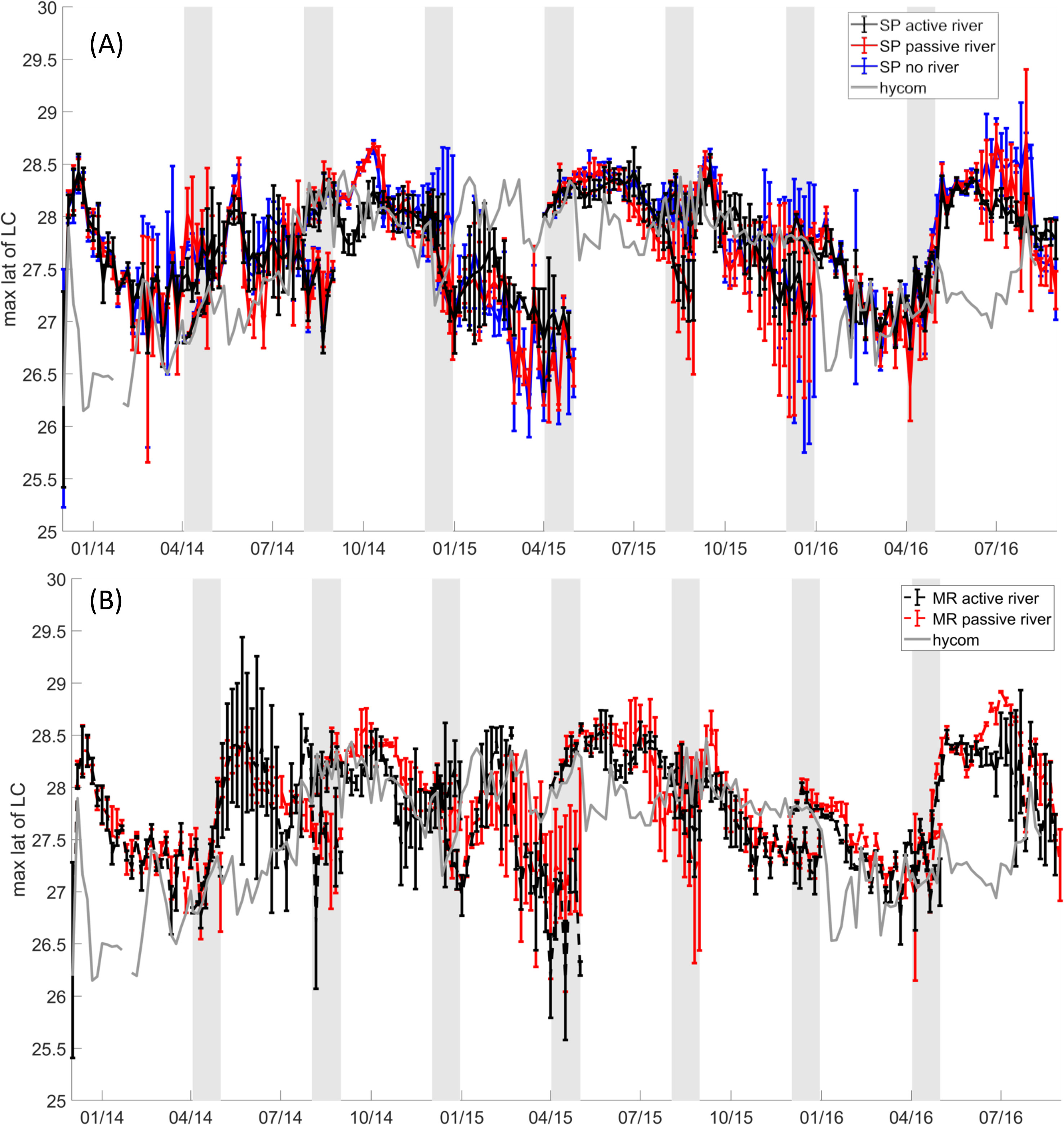}
\caption{Time-series of the northernmost latitude of LCS (LC or Rings) in the ensemble means of the different SP (A) and MR (B) configurations. The error bars indicate 1 standard deviation (STD) of the ensemble spread within the same configuration. The grey shadings indicates when the new integration period starts, overlapping with the previous period.}
\label{fig:LC_ext}
\end{figure}

Differences in the evolution of the northern extension of LC are therefore due to differences in the 1-day apart initial conditions and to the intrinsic chaoticity of the circulation in the northern portion of the basin. An example of these differences is shown in Fig. \ref{fig:SSH_LC}. We chose a time when CROCO and HYCOM are significantly different also to elucidate the case of Rings detachment ahead of what verified in the observation-based analysis. In this case, both the LC position, and the northward extension of the 17 cm SSHa contour differs among the ensemble members. In the three runs, the previously extended LC has just shed a Ring which was pushed quickly westward by the winds and has its core centered at 90$^o$W, but the shape and orientation of the Ring differ among the three realizations. The overall evolution, however, is comparable among members and most importantly among different ensembles, or in other words, the evolution of the ensemble mean is very similar among the ensembles in all 8 time periods. 
Indeed, in all ensembles, the LC and major Rings evolves similarly in CROCO (Fig. \ref{fig:LC_ext}), and most of the time their behavior compares well with that identified in the HYCOM analysis where all available satellite and in-situ observations are assimilated. This points to the relevance of the external forcing - the flow at Yucatan Strait and the wind stress - on the LC.  

Even though the SSH patterns appear similar, the exact location and northward extension of the LCS varies among runs. 
Differences among our ensembles and HYCOM are generally found when the Rings detach from LC, for example in May 2014 and in February (Fig. \ref{fig:SSH_LC}) and June 2015, or when two Rings coexist in CROCO, which occurs in our simulations during winter 2014 and spring 2015.
Unsurprisingly, the predictability potential is lowest at the time of Rings' detachment. 
Additionally, the comparison of the mesoscale patterns in the various runs suggests that the Rings' evolution is less predictable than that of the LC (i.e. the spread among ensembles is largest when the mesoscale circulation north of 24\textdegree N is dominated by one or more large Rings slowly moving westward). This is realized also in the observations and a quantification of it can be found in \cite{liu2021GRL}. 

Our resolution exploration shows that the spread is on average slightly larger in the SP ensembles compared to the MR cases, but the differences are not robust or consistent across the whole period. The STD among ensemble members are 0.16${^o}$ for SP active river, 0.21${^o}$ for SP passive river, 0.19${^o}$ for MR active river and 0.17${^o}$ for MR passive river.  We will further investigate this point while discussing surface vorticity.
The choice of river configuration, passive or active, on the other hand, does not modify the LCS pathway in a robust way, except for when the LC or the Rings extend into the northern portion of the basin, north of 28${^o}$N. In this extended condition the integrations with an active representation of the rivers limit the northward LCS reach. This is quantified in Table \ref{table:LC_count} that presents the probability that a given configuration simulates the northernmost extension of the LCS whenever the latitude furthest to the north achieved by the 17 cm SSH contour is north of of 28\textdegree N.  In this calculation it is important to notice that the least realistic no-river case is under-sampled compared to the others  and that we consider both the LC and the Rings. We use 5-day averaged SSH fields and find 90 events over a total of 239 for which the northern boundary of the LCS is found to the north of 28\textdegree N.

\begin{table}
\centering
\caption{(Northernmost reach of the LCS.}
\label{table:LC_count}
\centering
\begin{tabular}{l c c }
\hline
 Ensemble &  $>$ 28${^o}$N  \%\\
\hline
SP Active river &   2.2  \\
SP Passive river &  8.9 \\
SP No river & 26.7 \\
MR Active River & 13.3 \\
MR Passive River & 48.9 \\
\hline
\end{tabular}
\end{table}

Under these conditions, the northernmost LC is simulated by the MR passive river ensemble in nearly 50\% of the cases, followed by the SP no-river case (26.7 \%). The SP active river members accounts for only 2.2\% of the cases. We tested the significance of highest (MR passive) and lowest (SP active) probabilities using Monte Carlo simulations under the null hypothesis that the target event does not have the highest/lowest chance.  For the MR passive and SP active cases, the null hypothesis can be rejected with 99\%  and 95\% confidence level, respectively. Incidentally, if the LC 17 cm SSH contour does not exceed 28\textdegree N, the MR runs (active and passive combined) account for about 60\% of the remaining 149 events, and the SP simulations (active, passive and no-river) for approximately 40\%. Given the larger number of SP simulations, the predominance of MR cases is significant according to a binomial test at the 95\% confidence level. If the same calculations are repeated using 27.5\textdegree N as threshold instead, the highest chance significance of the MR passive case remains unchanged, but the null hypothesis cannot be rejected with 90\% confidence for the lowest chance (SP case). The predominance of combined MR cases, on the other hand, is significant at the  99\% confidence level.

\begin{figure}
\includegraphics[width=1\linewidth]{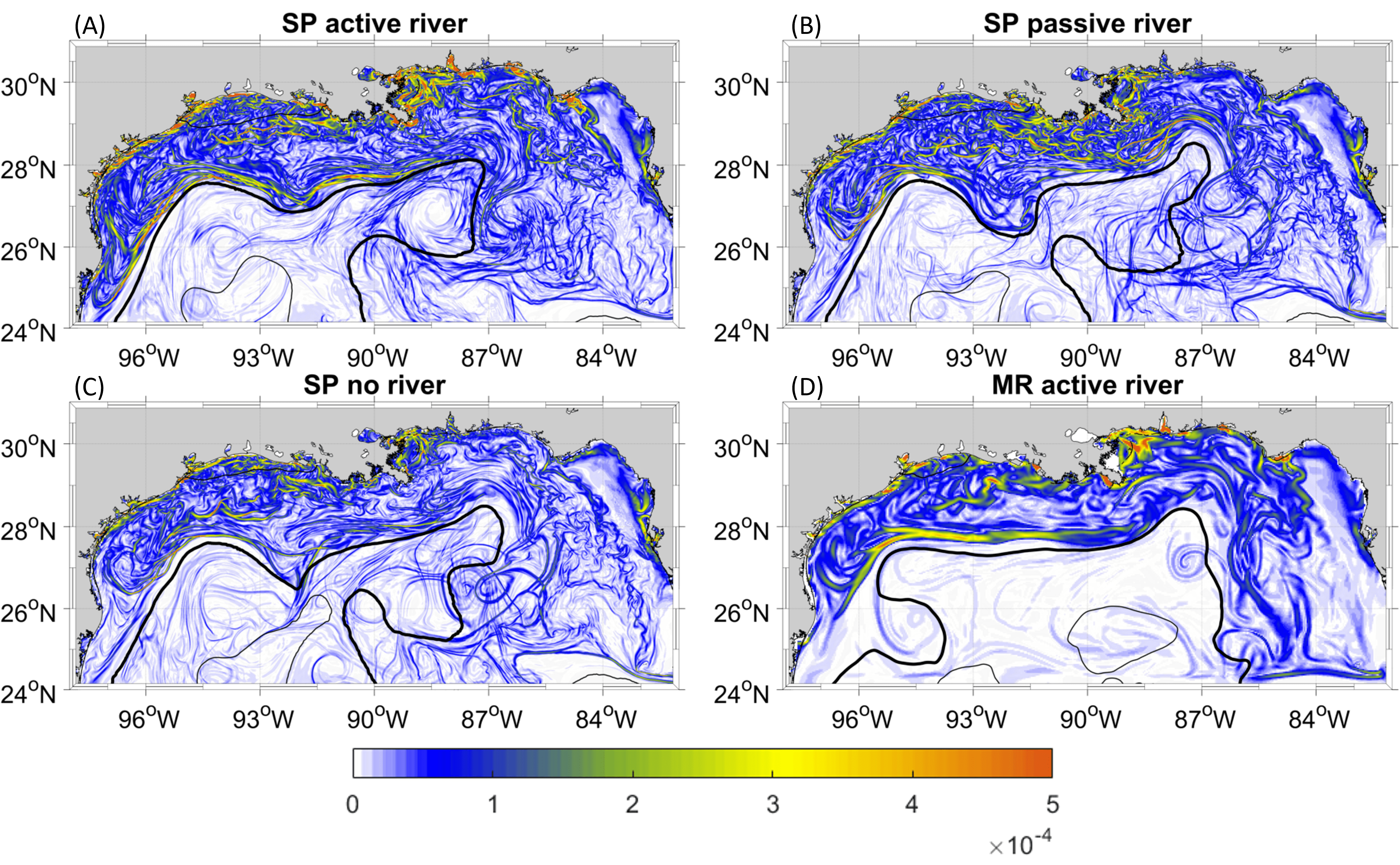}
\caption{Gradient of ensemble mean SSS in different configurations on Jul. 6, 2016. The black lines follow the 17 cm SSHa contours and the thick line identifies the LCS location.}
\label{fig:Sgrad_LC}
\end{figure}

What is limiting the LCS northward propagation in the SP runs with an active river? The LCS boundary especially in the summer seasons of 2015 and 2016 and in the winter of 2016, when the freshwater forcing was strong, is always outlined by intense salinity fronts. These fronts are  better resolved and more intense in the SP active-river runs, in which the submesoscale circulations, the more localized freshwater sources and the southward velocity associated with these riverine fluxes prevent the LC from extending towards the coast. 
An example is provided in Fig. \ref{fig:Sgrad_LC}. Noticeably the LCS boundary is not only located further south in the SP runs and more so in the active river configuration, but the salinity gradients extend into the LCS in the SP runs, indicating that a small amount of low salinity waters can penetrate the mesoscale transport barrier if submesoscale dynamics are resolved, eroding it. 
As counterexample, in the summer of 2014, following a season of below normal riverine fluxes into the basin, the penetration of the LCS north of 28\textdegree N is nearly indistinguishable among the SP runs, while still reaching its maximum extension in one of the MR integrations (not shown). 

The reason beyond the differences among MR and SP integrations resides in the submesoscale-induced mixing  at the LCS boundary, which is present at all times in the SP cases being the Loop Current and the Rings characterized by different water densities than the rest of the GoM. This mixing is strengthened in the northern portion of the basin by the riverine input, but exists independently of it to a lesser degree. The mixing is pertinent not only to salinity, but also to temperature, nutrients or pollutants, is missed entirely if the model resolution is too coarse to capture the submesoscale dynamics, and explains the overall LCS behavior in the SP versus MR runs.

\subsection{Ensemble spread of surface fields}
\label{sec:Results:spread}
\begin{figure}
\includegraphics[width=1\linewidth]{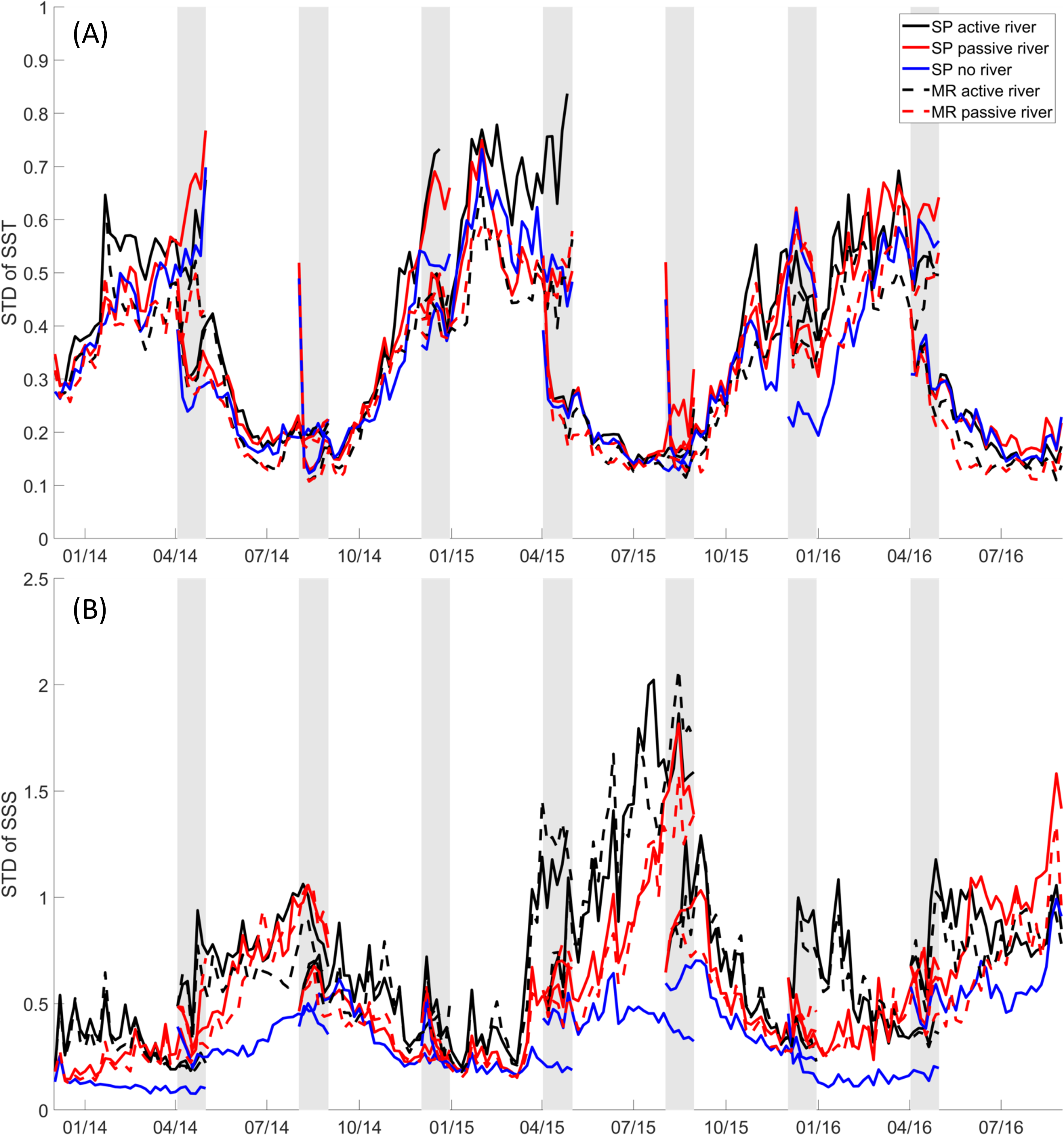}
\caption{Evolution of average STDs of SST (A) and SSS (B) among members with the same resolution and river configuration. The STDs are averaged over all ocean grid north of 27\textdegree N.}
\label{fig:STD_TS}
\end{figure}

\begin{figure}
\includegraphics[width=1\linewidth]{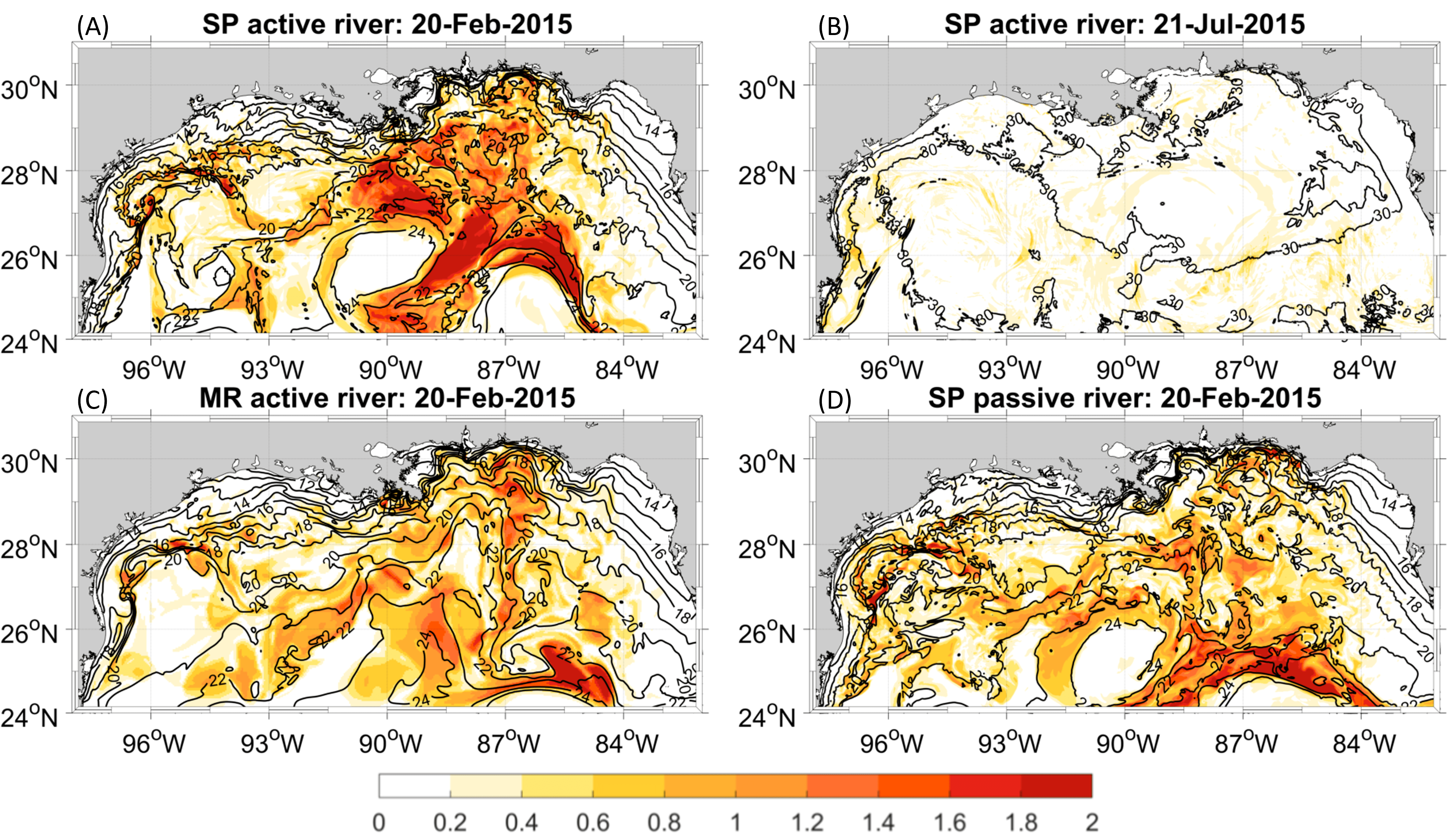}
\caption{STDs of SST among members of the SP active river runs (A-B), MR active river runs (C) and SP passive river runs (D) in winter 2015 (ACD) and summer 2015 (B).Contour lines indicate the ensemble mean SST at 1\textdegree C intervals.}
\label{fig:STD_T_patt}
\end{figure}

We next consider the evolution and patterns of ensemble spread in SST, SSS and surface vorticity. Fig. \ref{fig:STD_TS}A shows the evolution of the STD in SST. The initial SST differences are generally small except for runs starting in August, independently of the year. In summer the initial STD is large because the mixed layer is shallow everywhere in the basin, the heat fluxes can change significantly from one day to the next, and the resulting SST variability in the initial conditions can be large. 
The SST spread evolution over time shows clear seasonality, increasing in all years during fall and winter, with maximum spreading being achieved between February and April, and decreasing in summer from April to September with minima  between July and September (Fig. \ref{fig:STD_TS}A), when the atmospheric heat fluxes are strongest and winds generally weak.
In February, when the STD is at its maximum, the largest spread among ensemble members is near the edge of LC (Fig. \ref{fig:STD_T_patt}A), where the SST gradient is large, and small changes in the LC location and circulation among runs induce large local differences in the resulting ocean temperatures.
The SST gradient near the coast is also strong in winter: on the wide GoM shelves the shallow depths allow for large SST variations  among runs mainly controlled by mesoscale and smaller turbulence that extend their influence to the botttom due to the relatively strong winds and heat fluxes. In July-September, on the other hand, the surface mixed layer is shallow also over deep waters, the mean momentum and heat fluxes are weak, and the SST gradient is small over the whole basin, so the differences among the members and among ensembles are far more contained (see the STD on July 21st, 2015 in Fig. \ref{fig:STD_T_patt}B). In the summer season the surface temperature dynamics are dominated by the external forcing, the heat fluxes in particular, and is therefore potentially more predictable. 
The SST spread increases by increasing resolution, as turbulent structures at smaller scales are better resolved and chaotic, and this is  evident in winter in all years (compare Fig. \ref{fig:STD_T_patt}A and C). Between January and the end of March there is also a tendency for larger STD over the northern shelves in the active river configuration compared to the passive or no-river cases (see Fig. \ref{fig:STD_T_patt}A and D). This signal, quantified by the mean STD calculated for each active/passive ensemble in all periods (Table \ref{table:meanSTD_SST}) is however smaller than the resolution dependency.

\begin{table}
\centering
\caption{(Mean STD of SST in each ensemble in the 8 simulated periods.}
\label{table:meanSTD_SST}
\centering
\begin{tabular}{l c c c c}
\hline
Period & SP active & SP passive & MR active & MR passive \\
\hline
Dec. 2013 - Apr. 2014 & 0.49 & 0.47 & 0.41 & 0.38 \\
Apr. 2014 - Aug. 2014 & 0.26 & 0.25 & 0.24 & 0.22 \\
Aug. 2014 - Dec. 2014 & 0.34 & 0.35 & 0.29 & 0.29 \\
Dec. 2014 - Apr. 2015 & 0.63 & 0.51 & 0.47 & 0.49 \\
Apr. 2015 - Aug. 2015 & 0.21 & 0.22 & 0.19 & 0.18 \\
Aug. 2015 - Dec. 2015 & 0.36 & 0.36 & 0.30 & 0.35 \\
Dec. 2015 - Apr. 2016 & 0.52 & 0.53 & 0.46 & 0.45 \\
Apr. 2016 - Aug. 2016 & 0.22 & 0.23 & 0.19 & 0.18 \\
\hline
\end{tabular}
\end{table}

\begin{figure}
\includegraphics[width=1\linewidth]{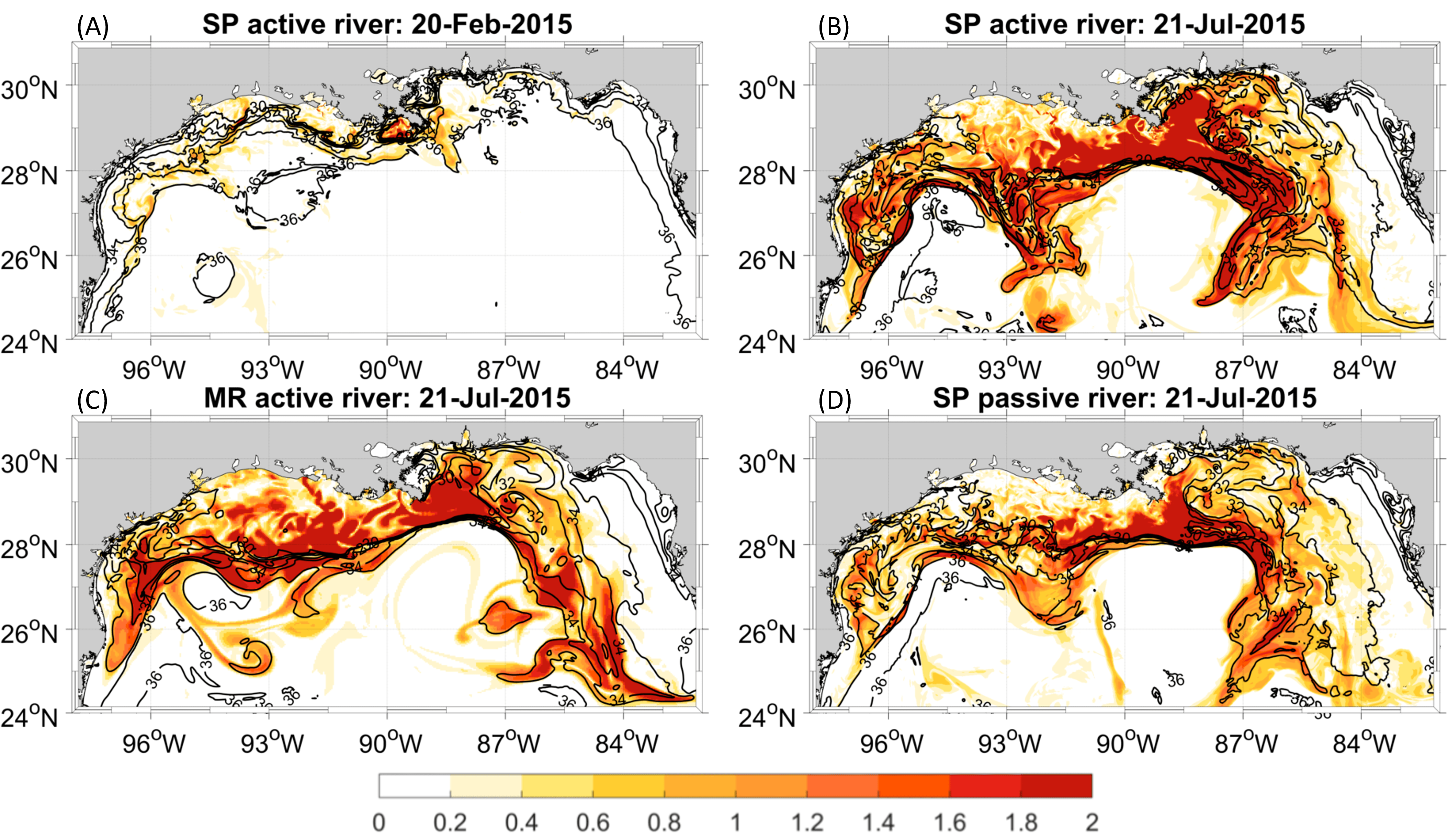}
\caption{STDs of SSS among members of the SP active river runs (A-B), MR active river runs (C) and SP passive river runs (D) in winter 2015 (A) and summer 2015 (B-C-D). Contour lines indicate the ensemble mean SSS at 1 psu interval (contour lines below 30 psu are not shown for clarity).}
\label{fig:STD_S_patt}
\end{figure}

The SSS spread has opposite seasonality and is greater than that of SST. It is generally contained in fall and winter (Fig. \ref{fig:STD_TS}B) and increases most rapidly in summer, from April to August, when the offshore spreading of riverine water into the GoM is at its maximum and the surface stratification is strong, inhibiting the vertical mixing of the freshwater contribution.
The SSS interannual spread is also larger than for SSTs, and follows the variations in river discharge.
In spring/summer 2014 for example, the river discharge was below average, and the STD of SSS is significantly smaller than in the same season of 2015, with an increased predictability potential. Following the 2016 winter flooding, on the other hand, a large STD is found also in the cold months in the active river ensemble.
In winter, the largest variance is found near the coast, over the wide shelves and the continental slope, where the SSS gradients are elevated (Fig. \ref{fig:STD_S_patt}A), while in summer the interaction of the mesoscale circulations with the freshwater determine the STD pattern, and the variance is particularly elevated whenever the LC or the Rings reach into the northern half of the basin \cite{schiller2014loop}.
The spread is greater among runs with an active river configuration since the fresh water enters the ocean over few grid points creating stronger gradients than in the passive river cases, in which  the discharge is distributed over a larger area. Despite the amplification of submesoscale fronts by the freshwater forcing in summer, the resolution dependency is relatively contained (Table \ref{table:meanSTD_SSS}) and smaller than found for SSTs. In other words, the model resolution has a small influence on the predictability of salinity anomalies, that are externally controlled by the amount of riverine input and winds, and internally by the mesoscale circulations with the LC and the Rings being the major players.

\begin{figure}
\includegraphics[width=1\linewidth]{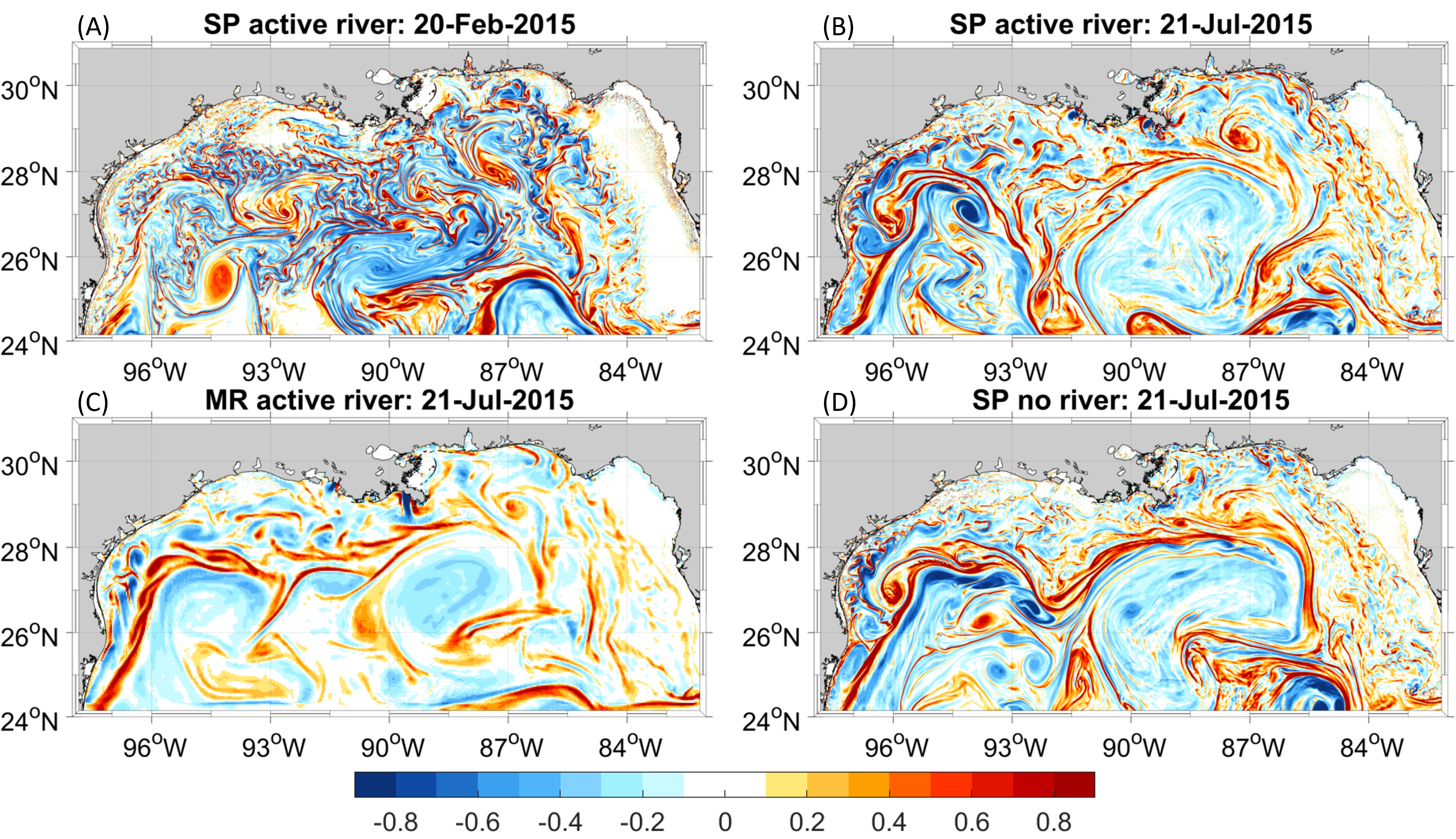}
\caption{Snapshot of surface vorticity normalized by Coriolis from the SP active river runs (A-B), MR active river runs (C) and SP no-river runs (D) in winter 2015 (A) and summer 2015 (B-C-D).}
\label{fig:surf_curl}
\end{figure}

\begin{table}
\centering
\caption{(Mean STD of SSS in each ensemble in the 8 simulated periods.}
\label{table:meanSTD_SSS}
\centering
\begin{tabular}{l c c c c}
\hline
Period & SP active & SP passive & MR active & MR passive \\
\hline
Dec. 2013 - Apr. 2014 & 0.32 & 0.27 & 0.29 & 0.24 \\
Apr. 2014 - Aug. 2014 & 0.78 & 0.69 & 0.65 & 0.65 \\
Aug. 2014 - Dec. 2014 & 0.54 & 0.41 & 0.49 & 0.38 \\
Dec. 2014 - Apr. 2015 & 0.57 & 0.37 & 0.59 & 0.35 \\
Apr. 2015 - Aug. 2015 & 1.25 & 0.94 & 1.27 & 0.84 \\
Aug. 2015 - Dec. 2015 & 0.65 & 0.58 & 0.64 & 0.53 \\
Dec. 2015 - Apr. 2016 & 0.57 & 0.45 & 0.54 & 0.42 \\
Apr. 2016 - Aug. 2016 & 0.86 & 0.90 & 0.79 & 0.68 \\
\hline
\end{tabular}
\end{table}

\begin{figure}
\includegraphics[width=1\linewidth]{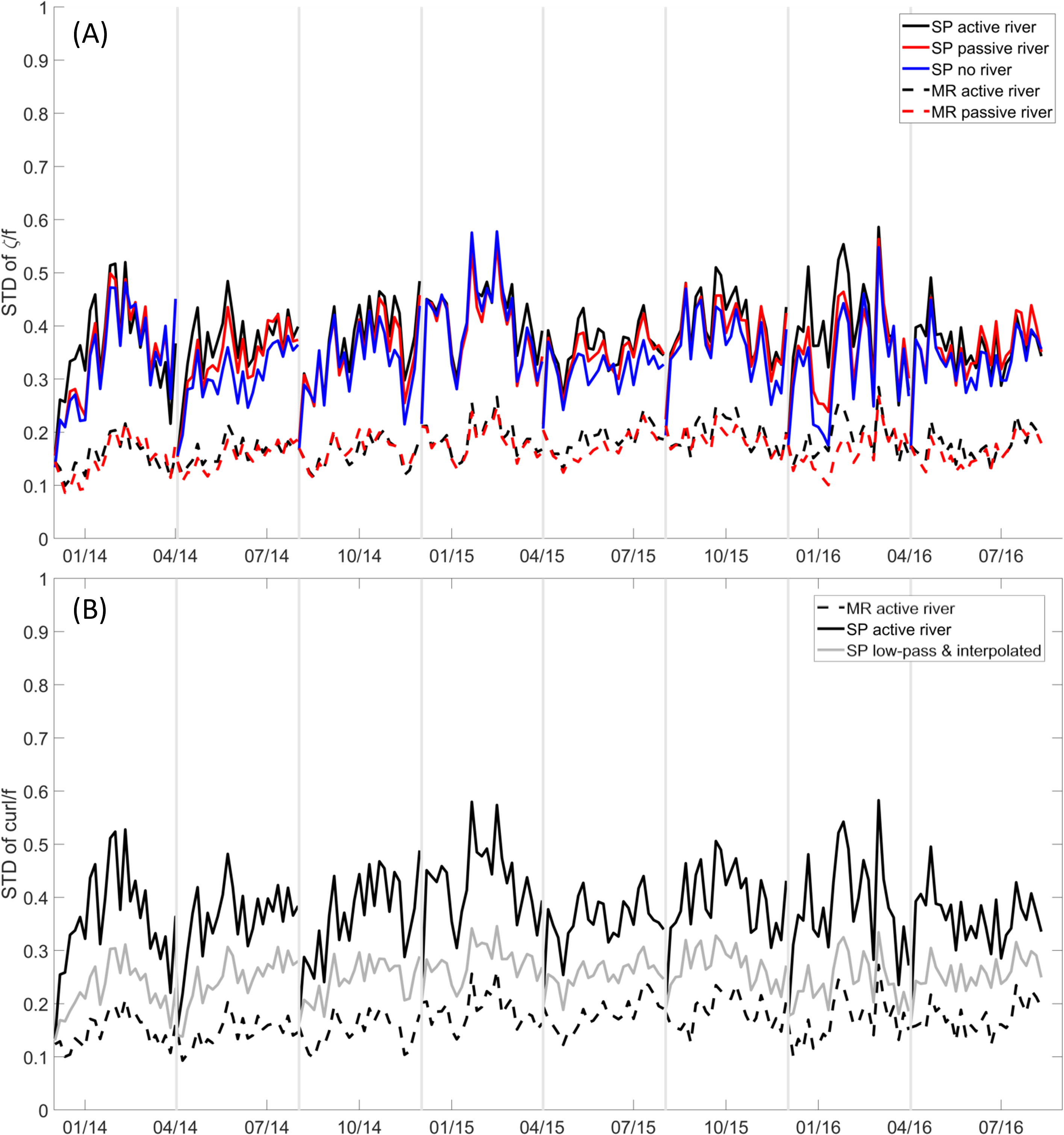}
\caption{Evolution of average STDs of surface vorticity (normalized by Coriolis) among members with the same resolution and river configuration (A), and comparison between STDs among members of MR active river runs (black dash curve), SP active river runs (black solid curve), and low-pass filtered and interpolated SP active river runs (magenta solid curve) (B).}
\label{fig:STD_curl}
\end{figure}

Finally, we analyze the spread in surface vorticity nondimensionalized by the Coriolis parameter and defined as $\zeta/f = 1/f(\frac{\partial v}{\partial x}-\frac{\partial u}{\partial y})$, where u and v are the zonal and meridional velocities (Fig. \ref{fig:surf_curl}). In this case, we find a strong resolution dependency, as to be expected.
The submesoscale circulations  are more intense in winter, when the mixed layer is 80-100 m deep, and relatively weaker in summer, when they are confined near the coast in correspondence of the strong density gradient created by the riverine input (Fig. \ref{fig:surf_curl}A - B)  \cite{luo2016submesoscale,barkan2017submesoscale1}.
Runs at SP resolution are characterized by a larger spread among ensemble members than those resolving only the mesoscales (Fig. \ref{fig:STD_curl}A), supporting the conclusions by \citeA{sandery2017ocean} that predictability decreases for increasing resolution of the ocean models.
The STD is also larger in winter than in other seasons, and is slightly amplified in the ensembles that include the riverine forcing, especially in summer (compare Fig. \ref{fig:surf_curl}A and D). As a result, the no-river ensemble shows lower values than either the active or passive river configurations at all times but for winter 2015, when the freshwater input was very low and did not reach offshore the continental shelf in any integration.

Figure \ref{fig:STD_curl}A reveals also that despite the difference in amplitude and the lack of seasonality in the STD of the lower resolution ensembles, the time-series corresponding to the SP and MR cases are highly correlated (the correlation coefficient is 0.68. This suggests that the model resolution controls the STD amplitude, but the STD time evolution depends on the representation of the larger mesoscale circulations and, in the GoM, on the LCS.  
We ask therefore if filtering the submesoscale circulations in the SP ensembles is sufficient to reduce the surface vorticity STD at the level of the corresponding 3.5 km ensembles, or if nonlinear interactions between the submesoscale and mesoscale circulations modify and limit the mesoscale predictability by impacting the modeled evolution of the latter. To answer this question, the SP resolution runs in the active river configuration are filtered using a low-pass Buttworth filter  to exclude the high frequency signal unresolved by the MP runs, and are then interpolated onto the 3.5 km horizontal resolution grid.
The STD of the resulting fields still shows larger spread than the MP runs (Fig. \ref{fig:STD_curl}B). This analysis  indicates that the SP runs include interactions between the submesoscale and mesoscale circulations that affect the evolution of the LCS and decrease its predictability potential.

\section{Conclusions}
\label{sec:Discussion}
In this work, we explored the predictability of surface fields in the northern Gulf of Mexico using ensembles of numerical integrations performed with a regional ocean model covering 2014, 2015 and the first 8 months of 2016. The model, CROCO, extends only to 24$^o$N and is forced at the open boundaries by the widely used HYCOM data assimilative analysis. Horizontal resolution (3.5 or 1 km) and representation of the riverine inflow (bogus and limited to the surface over an area extending offshore around the river mouth, or including a meridional momentum flux and decaying exponentially with depth at the river mouth) varied across the ensembles. Only the initial conditions differed by one day across members (up to three) in the same ensemble, and each integration lasted 5 months. 

Overall, under the conditions explored by this work, the surface ocean circulation predictability is high in the northern GoM if the atmospheric forcing is known. All members and all ensembles simulate a similar LCS behavior up to 5 months. CROCO - the regional model of choice - tend to overestimate the detachment of LC eddies compared to the analysis, but does so consistently across integrations. This result, taken together with findings by \citeA{cardona2016predictability} that argued for limited forecast skill in a comparable model that extended further to the South, supports the notion of an important role for the flow at the Yucatan Channel as forcing mechanism of the LC dynamics. Such outcome is in agreement with previous numerical \cite{ezer2003variability,oey1996simulation,oey2004vorticity,change2011loop,le2012surface} and observational studies \cite{Candela2002potential}.

We focused our attention on the interactions between the Loop Current system and the Mississippi-Achafalaya River System. While \citeA{schiller2011dynamics} have shown that offshore transport associated with the interaction of the river plume and the mesoscale circulations happens frequently and in all seasons, and depends on the mesoscale variability and on topographic and wind effects, here we asked how predictable are the surface fields affected by this transport.
By analyzing the sea surface heigh fields and the LCS position, we have shown that the LCS-MARS interaction is two-ways, with the LCS system being influenced by - and not only influencing - the freshwater plume. If the freshwater flux is strong, the northward extension of the LCS is constrained by the salinity gradients. In the regional model this influence is stronger if the riverine inflow is simulated in an active fashion with a velocity component proportional to the flux, and not like a bogus freshwater flux at the ocean surface, and increases with model resolution as frontogenetic instabilities are more prominent in submesoscale permitting simulations. 

In terms of sea surface temperature and salinity, the first shows less spread than the latter, unsurprisingly. However, the two variables have opposite seasonality in their predictability signal. In summer, when the mixed layer is shallow, the SST field shows a relatively small spread among members and among ensembles, being their variability driven by the large-scale atmospheric heat fluxes. In winter, the mesoscale variability and the temperature gradients associated with the LCS in offshore waters and with turbulence over the shelf decrease the predictability potential. For surface salinity, instead, predictability is inversely proportional to the river discharge and is usually reduced in summer, especially around the LCS and over the shelf. As a result of its discharge dependency, the SSS predictability is also affected by the interannual variations in riverine input.
For both SST and SSS increasing resolution and accuracy in the representation of the riverine fluxes causes the spread to increase. This is more evident for temperature than for salinity. We plan to  further investigate how the configuration changes modify the propagation of freshwater anomalies throughout the water column in a separate study. 

Finally, we have shown that the interactions of the mesoscale and submesoscale circulations result in nonlinear feedbacks from the smaller to the larger scales. The predictability of the surface vorticity changes in amplitude when increasing resolution but retains most of its intraseasonal and interannual signal. However, submesoscale simulations downscaled to mesoscale resolving resolution continue to have greater spread than the MR ensembles.
This is consistent with results by \citeA{sandery2017ocean}.

Ensemble members differ only in their initial conditions. Even though this is a key source of uncertainty in predicting the evolution of any flow field, other perturbations, such as those at the open boundaries or in model parameters or atmospheric forcing fields, could and should be investigated in future research. How uncertainty in the atmospheric forcing translates in a degeneration of ocean predictability is particularly important as intrinsic to any ocean forecast. Additionally, we note that from a theoretical standpoint, \citeA{barker1991relationship} has shown that the correlation between the ensemble model spread and the forecast skill may be weak at long forecast times (several times the decorrelation time scale of the flow under investigation), so the in the Gulf of Mexico our upper bound evaluation of the predictability potential may not be reliable for more than 2-3 months.

\acknowledgments
This work was supported entirely by the Gulf Research Program (GRP) of the National Academies of Sciences, Engineering, and Medicine through its Understanding Gulf Ocean Systems 2 (UGOS-2) Grant “The Loop Current and the Mississippi-Atchafalaya River System: Interactions, Variability and Modeling Requirements”.
All model runs used in this research are made available via Dropbox link, \url{https://www.dropbox.com/sh/5iycwzkybjr3rl5/AAC0zv9T0bn064j3__gwv7Vxa?dl=0}. The Georgia Institute of Technology has a campus-wide license for Dropbox, providing unlimited, ITAR-compliant cloud storage space for large modeling data-sets. CROCO is a community model. The version used in this work is 1.0 and can be found at \url{https://www.croco-ocean.org/download/croco-project/}. The NAVGEM data-set used to force CROCO was downloaded from: \url{https://www.hycom.org/dataserver/navgem} (last access: 08/17/2021).
The HYCOM analysis data can be accessed through the \url{https://www.hycom.org/dataserver} (\url{https://www.hycom.org/data/goml0pt04/expt-31pt0}, and \url{https://www.hycom.org/data/goml0pt04/expt-32pt5}) (last access: 08/17/2021). Codes for figure plotting can be found at \url{https://doi.org/10.5281/zenodo.5504606} (DOI: 10.5281/zenodo.5504606).


%
%

\bibliography{GOMreferences}

\begin{thebibliography}{}

\bibitem [\protect \citeauthoryear {%
Androulidakis%
\ \protect \BOthers {.}}{%
Androulidakis%
\ \protect \BOthers {.}}{%
{\protect \APACyear {2019}}%
}]{%
androulidakis2019offshore}
\APACinsertmetastar {%
androulidakis2019offshore}%
\begin{APACrefauthors}%
Androulidakis, Y.%
, Kourafalou, V.%
, Le~H{\'e}naff, M.%
, Kang, H.%
, Sutton, T.%
, Chen, S.%
\BDBL {}Ntaganou, N.%
\end{APACrefauthors}%
\unskip\
\newblock
\APACrefYearMonthDay{2019}{}{}.
\newblock
{\BBOQ}\APACrefatitle {Offshore spreading of Mississippi waters: Pathways and
  vertical structure under eddy influence} {Offshore spreading of mississippi
  waters: Pathways and vertical structure under eddy influence}.{\BBCQ}
\newblock
\APACjournalVolNumPages{Journal of Geophysical Research:
  Oceans}{124}{8}{5952--5978}.
\PrintBackRefs{\CurrentBib}

\bibitem [\protect \citeauthoryear {%
Androulidakis%
\ \protect \BOthers {.}}{%
Androulidakis%
\ \protect \BOthers {.}}{%
{\protect \APACyear {2018}}%
}]{%
androulidakis2018influence}
\APACinsertmetastar {%
androulidakis2018influence}%
\begin{APACrefauthors}%
Androulidakis, Y.%
, Kourafalou, V.%
, {\"O}zg{\"o}kmen, T.%
, Garcia-Pineda, O.%
, Lund, B.%
, Le~H{\'e}naff, M.%
\BDBL {}others%
\end{APACrefauthors}%
\unskip\
\newblock
\APACrefYearMonthDay{2018}{}{}.
\newblock
{\BBOQ}\APACrefatitle {Influence of river-induced fronts on hydrocarbon
  transport: A multiplatform observational study} {Influence of river-induced
  fronts on hydrocarbon transport: A multiplatform observational study}.{\BBCQ}
\newblock
\APACjournalVolNumPages{Journal of Geophysical Research:
  Oceans}{123}{5}{3259--3285}.
\PrintBackRefs{\CurrentBib}

\bibitem [\protect \citeauthoryear {%
Auclair%
\ \protect \BOthers {.}}{%
Auclair%
\ \protect \BOthers {.}}{%
{\protect \APACyear {2018}}%
}]{%
auclair2018some}
\APACinsertmetastar {%
auclair2018some}%
\begin{APACrefauthors}%
Auclair, F.%
, Benshila, R.%
, Debreu, L.%
, Ducousso, N.%
, Dumas, F.%
, Marchesiello, P.%
\BCBL {}\ \BBA {} Lemari{\'e}, F.%
\end{APACrefauthors}%
\unskip\
\newblock
\APACrefYearMonthDay{2018}{}{}.
\newblock
{\BBOQ}\APACrefatitle {Some recent developments around the CROCO initiative for
  complex regional to coastal modeling} {Some recent developments around the
  croco initiative for complex regional to coastal modeling}.{\BBCQ}
\newblock
\BIn{} \APACrefbtitle {COMOD 2018-Workshop on Coastal Ocean Modelling} {Comod
  2018-workshop on coastal ocean modelling}\ (\BPGS\ 1--47).
\PrintBackRefs{\CurrentBib}

\bibitem [\protect \citeauthoryear {%
Barkan%
, McWilliams%
, Molemaker%
\BCBL {}\ \protect \BOthers {.}}{%
Barkan%
, McWilliams%
, Molemaker%
\BCBL {}\ \protect \BOthers {.}}{%
{\protect \APACyear {2017}}%
}]{%
barkan2017submesoscale2}
\APACinsertmetastar {%
barkan2017submesoscale2}%
\begin{APACrefauthors}%
Barkan, R.%
, McWilliams, J\BPBI C.%
, Molemaker, M\BPBI J.%
, Choi, J.%
, Srinivasan, K.%
, Shchepetkin, A\BPBI F.%
\BCBL {}\ \BBA {} Bracco, A.%
\end{APACrefauthors}%
\unskip\
\newblock
\APACrefYearMonthDay{2017}{}{}.
\newblock
{\BBOQ}\APACrefatitle {Submesoscale dynamics in the northern Gulf of Mexico.
  Part II: Temperature--salinity relations and cross-shelf transport processes}
  {Submesoscale dynamics in the northern gulf of mexico. part ii:
  Temperature--salinity relations and cross-shelf transport processes}.{\BBCQ}
\newblock
\APACjournalVolNumPages{Journal of Physical Oceanography}{47}{9}{2347--2360}.
\PrintBackRefs{\CurrentBib}

\bibitem [\protect \citeauthoryear {%
Barkan%
, McWilliams%
, Shchepetkin%
\BCBL {}\ \protect \BOthers {.}}{%
Barkan%
, McWilliams%
, Shchepetkin%
\BCBL {}\ \protect \BOthers {.}}{%
{\protect \APACyear {2017}}%
}]{%
barkan2017submesoscale1}
\APACinsertmetastar {%
barkan2017submesoscale1}%
\begin{APACrefauthors}%
Barkan, R.%
, McWilliams, J\BPBI C.%
, Shchepetkin, A\BPBI F.%
, Molemaker, M\BPBI J.%
, Renault, L.%
, Bracco, A.%
\BCBL {}\ \BBA {} Choi, J.%
\end{APACrefauthors}%
\unskip\
\newblock
\APACrefYearMonthDay{2017}{}{}.
\newblock
{\BBOQ}\APACrefatitle {Submesoscale dynamics in the northern Gulf of Mexico.
  Part I: Regional and seasonal characterization and the role of river outflow}
  {Submesoscale dynamics in the northern gulf of mexico. part i: Regional and
  seasonal characterization and the role of river outflow}.{\BBCQ}
\newblock
\APACjournalVolNumPages{Journal of Physical Oceanography}{47}{9}{2325--2346}.
\PrintBackRefs{\CurrentBib}

\bibitem [\protect \citeauthoryear {%
Barker%
}{%
Barker%
}{%
{\protect \APACyear {1991}}%
}]{%
barker1991relationship}
\APACinsertmetastar {%
barker1991relationship}%
\begin{APACrefauthors}%
Barker, T\BPBI W.%
\end{APACrefauthors}%
\unskip\
\newblock
\APACrefYearMonthDay{1991}{}{}.
\newblock
{\BBOQ}\APACrefatitle {The relationship between spread and forecast error in
  extended-range forecasts} {The relationship between spread and forecast error
  in extended-range forecasts}.{\BBCQ}
\newblock
\APACjournalVolNumPages{Journal of climate}{4}{7}{733--742}.
\PrintBackRefs{\CurrentBib}

\bibitem [\protect \citeauthoryear {%
Bracco%
, Liu%
\BCBL {}\ \BBA {} Sun%
}{%
Bracco%
\ \protect \BOthers {.}}{%
{\protect \APACyear {2019}}%
}]{%
bracco2019mesoscale}
\APACinsertmetastar {%
bracco2019mesoscale}%
\begin{APACrefauthors}%
Bracco, A.%
, Liu, G.%
\BCBL {}\ \BBA {} Sun, D.%
\end{APACrefauthors}%
\unskip\
\newblock
\APACrefYearMonthDay{2019}{}{}.
\newblock
{\BBOQ}\APACrefatitle {Mesoscale-submesoscale interactions in the Gulf of
  Mexico: From oil dispersion to climate} {Mesoscale-submesoscale interactions
  in the gulf of mexico: From oil dispersion to climate}.{\BBCQ}
\newblock
\APACjournalVolNumPages{Chaos, Solitons \& Fractals}{119}{}{63--72}.
\PrintBackRefs{\CurrentBib}

\bibitem [\protect \citeauthoryear {%
Buizza%
}{%
Buizza%
}{%
{\protect \APACyear {1997}}%
}]{%
buizza1997potential}
\APACinsertmetastar {%
buizza1997potential}%
\begin{APACrefauthors}%
Buizza, R.%
\end{APACrefauthors}%
\unskip\
\newblock
\APACrefYearMonthDay{1997}{}{}.
\newblock
{\BBOQ}\APACrefatitle {Potential forecast skill of ensemble prediction and
  spread and skill distributions of the ECMWF ensemble prediction system}
  {Potential forecast skill of ensemble prediction and spread and skill
  distributions of the ecmwf ensemble prediction system}.{\BBCQ}
\newblock
\APACjournalVolNumPages{Monthly Weather Review}{125}{1}{99--119}.
\PrintBackRefs{\CurrentBib}

\bibitem [\protect \citeauthoryear {%
Callies%
, Flierl%
, Ferrari%
\BCBL {}\ \BBA {} Fox-Kemper%
}{%
Callies%
\ \protect \BOthers {.}}{%
{\protect \APACyear {2016}}%
}]{%
callies2016role}
\APACinsertmetastar {%
callies2016role}%
\begin{APACrefauthors}%
Callies, J.%
, Flierl, G.%
, Ferrari, R.%
\BCBL {}\ \BBA {} Fox-Kemper, B.%
\end{APACrefauthors}%
\unskip\
\newblock
\APACrefYearMonthDay{2016}{}{}.
\newblock
{\BBOQ}\APACrefatitle {The role of mixed-layer instabilities in submesoscale
  turbulence} {The role of mixed-layer instabilities in submesoscale
  turbulence}.{\BBCQ}
\newblock
\APACjournalVolNumPages{Journal of Fluid Mechanics}{788}{}{5--41}.
\PrintBackRefs{\CurrentBib}

\bibitem [\protect \citeauthoryear {%
Candela%
, Sheinbaum%
, Ochoa%
, Badan%
\BCBL {}\ \BBA {} Leben%
}{%
Candela%
\ \protect \BOthers {.}}{%
{\protect \APACyear {{2002}}}%
}]{%
Candela2002potential}
\APACinsertmetastar {%
Candela2002potential}%
\begin{APACrefauthors}%
Candela, J.%
, Sheinbaum, J.%
, Ochoa, J.%
, Badan, A.%
\BCBL {}\ \BBA {} Leben, R.%
\end{APACrefauthors}%
\unskip\
\newblock
\APACrefYearMonthDay{{2002}}{}{}.
\newblock
{\BBOQ}\APACrefatitle {{The potential vorticity flux through the Yucatan
  Channel and the Loop Current in the Gulf of Mexico}} {{The potential
  vorticity flux through the Yucatan Channel and the Loop Current in the Gulf
  of Mexico}}.{\BBCQ}
\newblock
\APACjournalVolNumPages{Geophysical Research Letters}{{29}}{{22}}{}.
\PrintBackRefs{\CurrentBib}

\bibitem [\protect \citeauthoryear {%
Cardona%
\ \BBA {} Bracco%
}{%
Cardona%
\ \BBA {} Bracco%
}{%
{\protect \APACyear {2016}}%
}]{%
cardona2016predictability}
\APACinsertmetastar {%
cardona2016predictability}%
\begin{APACrefauthors}%
Cardona, Y.%
\BCBT {}\ \BBA {} Bracco, A.%
\end{APACrefauthors}%
\unskip\
\newblock
\APACrefYearMonthDay{2016}{}{}.
\newblock
{\BBOQ}\APACrefatitle {Predictability of mesoscale circulation throughout the
  water column in the Gulf of Mexico} {Predictability of mesoscale circulation
  throughout the water column in the gulf of mexico}.{\BBCQ}
\newblock
\APACjournalVolNumPages{Deep Sea Research Part II: Topical Studies in
  Oceanography}{129}{}{332--349}.
\PrintBackRefs{\CurrentBib}

\bibitem [\protect \citeauthoryear {%
Chang%
\ \BBA {} Oey%
}{%
Chang%
\ \BBA {} Oey%
}{%
{\protect \APACyear {{2011}}}%
}]{%
change2011loop}
\APACinsertmetastar {%
change2011loop}%
\begin{APACrefauthors}%
Chang, Y\BPBI L.%
\BCBT {}\ \BBA {} Oey, L\BPBI Y.%
\end{APACrefauthors}%
\unskip\
\newblock
\APACrefYearMonthDay{{2011}}{}{}.
\newblock
{\BBOQ}\APACrefatitle {{Loop Current Cycle: Coupled Response of the Loop
  Current with Deep Flows}} {{Loop Current Cycle: Coupled Response of the Loop
  Current with Deep Flows}}.{\BBCQ}
\newblock
\APACjournalVolNumPages{Journal of Physical
  Oceanography}{{41}}{{3}}{{458-471}}.
\PrintBackRefs{\CurrentBib}

\bibitem [\protect \citeauthoryear {%
Counillon%
\ \BBA {} Bertino%
}{%
Counillon%
\ \BBA {} Bertino%
}{%
{\protect \APACyear {2009}}%
}]{%
counillon2009high}
\APACinsertmetastar {%
counillon2009high}%
\begin{APACrefauthors}%
Counillon, F.%
\BCBT {}\ \BBA {} Bertino, L.%
\end{APACrefauthors}%
\unskip\
\newblock
\APACrefYearMonthDay{2009}{}{}.
\newblock
{\BBOQ}\APACrefatitle {High-resolution ensemble forecasting for the Gulf of
  Mexico eddies and fronts} {High-resolution ensemble forecasting for the gulf
  of mexico eddies and fronts}.{\BBCQ}
\newblock
\APACjournalVolNumPages{Ocean Dynamics}{59}{1}{83--95}.
\PrintBackRefs{\CurrentBib}

\bibitem [\protect \citeauthoryear {%
Cummings%
}{%
Cummings%
}{%
{\protect \APACyear {2005}}%
}]{%
cummings2005operational}
\APACinsertmetastar {%
cummings2005operational}%
\begin{APACrefauthors}%
Cummings, J\BPBI A.%
\end{APACrefauthors}%
\unskip\
\newblock
\APACrefYearMonthDay{2005}{}{}.
\newblock
{\BBOQ}\APACrefatitle {Operational multivariate ocean data assimilation}
  {Operational multivariate ocean data assimilation}.{\BBCQ}
\newblock
\APACjournalVolNumPages{Quarterly Journal of the Royal Meteorological Society:
  A journal of the atmospheric sciences, applied meteorology and physical
  oceanography}{131}{613}{3583--3604}.
\PrintBackRefs{\CurrentBib}

\bibitem [\protect \citeauthoryear {%
Cummings%
\ \BBA {} Smedstad%
}{%
Cummings%
\ \BBA {} Smedstad%
}{%
{\protect \APACyear {2013}}%
}]{%
cummings2013variational}
\APACinsertmetastar {%
cummings2013variational}%
\begin{APACrefauthors}%
Cummings, J\BPBI A.%
\BCBT {}\ \BBA {} Smedstad, O\BPBI M.%
\end{APACrefauthors}%
\unskip\
\newblock
\APACrefYearMonthDay{2013}{}{}.
\newblock
{\BBOQ}\APACrefatitle {Variational data assimilation for the global ocean}
  {Variational data assimilation for the global ocean}.{\BBCQ}
\newblock
\BIn{} \APACrefbtitle {Data Assimilation for Atmospheric, Oceanic and
  Hydrologic Applications (Vol. II)} {Data assimilation for atmospheric,
  oceanic and hydrologic applications (vol. ii)}\ (\BPGS\ 303--343).
\newblock
\APACaddressPublisher{}{Springer}.
\PrintBackRefs{\CurrentBib}

\bibitem [\protect \citeauthoryear {%
Di~Marco%
, Nowlin%
\BCBL {}\ \BBA {} Reid%
}{%
Di~Marco%
\ \protect \BOthers {.}}{%
{\protect \APACyear {2005}}%
}]{%
dimarco2005statistical}
\APACinsertmetastar {%
dimarco2005statistical}%
\begin{APACrefauthors}%
Di~Marco, S\BPBI F.%
, Nowlin, W\BPBI D.%
\BCBL {}\ \BBA {} Reid, R.%
\end{APACrefauthors}%
\unskip\
\newblock
\APACrefYearMonthDay{2005}{}{}.
\newblock
{\BBOQ}\APACrefatitle {A statistical description of the velocity fields from
  upper ocean drifters in the Gulf of Mexico} {A statistical description of the
  velocity fields from upper ocean drifters in the gulf of mexico}.{\BBCQ}
\newblock
\APACjournalVolNumPages{Geophysical Monograph-American Geophysical
  Union}{161}{}{101}.
\PrintBackRefs{\CurrentBib}

\bibitem [\protect \citeauthoryear {%
Donohue%
, Watts%
, Hamilton%
, Leben%
\BCBL {}\ \BBA {} Kennelly%
}{%
Donohue%
\ \protect \BOthers {.}}{%
{\protect \APACyear {2016}}%
}]{%
donohue2016loop}
\APACinsertmetastar {%
donohue2016loop}%
\begin{APACrefauthors}%
Donohue, K\BPBI A.%
, Watts, D.%
, Hamilton, P.%
, Leben, R.%
\BCBL {}\ \BBA {} Kennelly, M.%
\end{APACrefauthors}%
\unskip\
\newblock
\APACrefYearMonthDay{2016}{}{}.
\newblock
{\BBOQ}\APACrefatitle {Loop current eddy formation and baroclinic instability}
  {Loop current eddy formation and baroclinic instability}.{\BBCQ}
\newblock
\APACjournalVolNumPages{Dynamics of Atmospheres and Oceans}{76}{}{195--216}.
\PrintBackRefs{\CurrentBib}

\bibitem [\protect \citeauthoryear {%
Dunn%
}{%
Dunn%
}{%
{\protect \APACyear {1996}}%
}]{%
dunn1996trends}
\APACinsertmetastar {%
dunn1996trends}%
\begin{APACrefauthors}%
Dunn, D\BPBI D.%
\end{APACrefauthors}%
\unskip\
\newblock
\APACrefYear{1996}.
\newblock
\APACrefbtitle {Trends in nutrient inflows to the Gulf of Mexico from streams
  draining the conterminous United States, 1972-93} {Trends in nutrient inflows
  to the gulf of mexico from streams draining the conterminous united states,
  1972-93}\ (\BVOL~96)\ (\BNUM\ 4113).
\newblock
\APACaddressPublisher{}{US Department of the Interior, US Geological Survey}.
\PrintBackRefs{\CurrentBib}

\bibitem [\protect \citeauthoryear {%
D’Asaro%
\ \protect \BOthers {.}}{%
D’Asaro%
\ \protect \BOthers {.}}{%
{\protect \APACyear {2018}}%
}]{%
d2018ocean}
\APACinsertmetastar {%
d2018ocean}%
\begin{APACrefauthors}%
D’Asaro, E\BPBI A.%
, Shcherbina, A\BPBI Y.%
, Klymak, J\BPBI M.%
, Molemaker, J.%
, Novelli, G.%
, Guigand, C\BPBI M.%
\BDBL {}others%
\end{APACrefauthors}%
\unskip\
\newblock
\APACrefYearMonthDay{2018}{}{}.
\newblock
{\BBOQ}\APACrefatitle {Ocean convergence and the dispersion of flotsam} {Ocean
  convergence and the dispersion of flotsam}.{\BBCQ}
\newblock
\APACjournalVolNumPages{Proceedings of the National Academy of
  Sciences}{115}{6}{1162--1167}.
\PrintBackRefs{\CurrentBib}

\bibitem [\protect \citeauthoryear {%
Egbert%
\ \BBA {} Erofeeva%
}{%
Egbert%
\ \BBA {} Erofeeva%
}{%
{\protect \APACyear {2002}}%
}]{%
egbert2002efficient}
\APACinsertmetastar {%
egbert2002efficient}%
\begin{APACrefauthors}%
Egbert, G\BPBI D.%
\BCBT {}\ \BBA {} Erofeeva, S\BPBI Y.%
\end{APACrefauthors}%
\unskip\
\newblock
\APACrefYearMonthDay{2002}{}{}.
\newblock
{\BBOQ}\APACrefatitle {Efficient inverse modeling of barotropic ocean tides}
  {Efficient inverse modeling of barotropic ocean tides}.{\BBCQ}
\newblock
\APACjournalVolNumPages{Journal of Atmospheric and Oceanic
  technology}{19}{2}{183--204}.
\PrintBackRefs{\CurrentBib}

\bibitem [\protect \citeauthoryear {%
Ezer%
, Oey%
, Lee%
\BCBL {}\ \BBA {} Sturges%
}{%
Ezer%
\ \protect \BOthers {.}}{%
{\protect \APACyear {2003}}%
}]{%
ezer2003variability}
\APACinsertmetastar {%
ezer2003variability}%
\begin{APACrefauthors}%
Ezer, T.%
, Oey, L\BHBI Y.%
, Lee, H\BHBI C.%
\BCBL {}\ \BBA {} Sturges, W.%
\end{APACrefauthors}%
\unskip\
\newblock
\APACrefYearMonthDay{2003}{}{}.
\newblock
{\BBOQ}\APACrefatitle {The variability of currents in the Yucatan Channel:
  Analysis of results from a numerical ocean model} {The variability of
  currents in the yucatan channel: Analysis of results from a numerical ocean
  model}.{\BBCQ}
\newblock
\APACjournalVolNumPages{Journal of Geophysical Research: Oceans}{108}{C1}{}.
\PrintBackRefs{\CurrentBib}

\bibitem [\protect \citeauthoryear {%
Gopalakrishnan%
, Cornuelle%
, Hoteit%
, Rudnick%
\BCBL {}\ \BBA {} Owens%
}{%
Gopalakrishnan%
\ \protect \BOthers {.}}{%
{\protect \APACyear {2013}}%
}]{%
gopalakrishnan2013state}
\APACinsertmetastar {%
gopalakrishnan2013state}%
\begin{APACrefauthors}%
Gopalakrishnan, G.%
, Cornuelle, B\BPBI D.%
, Hoteit, I.%
, Rudnick, D\BPBI L.%
\BCBL {}\ \BBA {} Owens, W\BPBI B.%
\end{APACrefauthors}%
\unskip\
\newblock
\APACrefYearMonthDay{2013}{}{}.
\newblock
{\BBOQ}\APACrefatitle {State estimates and forecasts of the loop current in the
  Gulf of Mexico using the MITgcm and its adjoint} {State estimates and
  forecasts of the loop current in the gulf of mexico using the mitgcm and its
  adjoint}.{\BBCQ}
\newblock
\APACjournalVolNumPages{Journal of Geophysical Research:
  Oceans}{118}{7}{3292--3314}.
\PrintBackRefs{\CurrentBib}

\bibitem [\protect \citeauthoryear {%
Hamilton%
}{%
Hamilton%
}{%
{\protect \APACyear {1990}}%
}]{%
hamilton1990deep}
\APACinsertmetastar {%
hamilton1990deep}%
\begin{APACrefauthors}%
Hamilton, P.%
\end{APACrefauthors}%
\unskip\
\newblock
\APACrefYearMonthDay{1990}{}{}.
\newblock
{\BBOQ}\APACrefatitle {Deep currents in the Gulf of Mexico} {Deep currents in
  the gulf of mexico}.{\BBCQ}
\newblock
\APACjournalVolNumPages{Journal of Physical Oceanography}{20}{7}{1087--1104}.
\PrintBackRefs{\CurrentBib}

\bibitem [\protect \citeauthoryear {%
Hamilton%
}{%
Hamilton%
}{%
{\protect \APACyear {2009}}%
}]{%
hamilton2009topographic}
\APACinsertmetastar {%
hamilton2009topographic}%
\begin{APACrefauthors}%
Hamilton, P.%
\end{APACrefauthors}%
\unskip\
\newblock
\APACrefYearMonthDay{2009}{}{}.
\newblock
{\BBOQ}\APACrefatitle {Topographic Rossby waves in the Gulf of Mexico}
  {Topographic rossby waves in the gulf of mexico}.{\BBCQ}
\newblock
\APACjournalVolNumPages{Progress in Oceanography}{82}{1}{1--31}.
\PrintBackRefs{\CurrentBib}

\bibitem [\protect \citeauthoryear {%
Hamilton%
\ \BBA {} Lee%
}{%
Hamilton%
\ \BBA {} Lee%
}{%
{\protect \APACyear {2005}}%
}]{%
hamilton2005eddies}
\APACinsertmetastar {%
hamilton2005eddies}%
\begin{APACrefauthors}%
Hamilton, P.%
\BCBT {}\ \BBA {} Lee, T\BPBI N.%
\end{APACrefauthors}%
\unskip\
\newblock
\APACrefYearMonthDay{2005}{}{}.
\newblock
{\BBOQ}\APACrefatitle {Eddies and jets over the slope of the northeast Gulf of
  Mexico} {Eddies and jets over the slope of the northeast gulf of
  mexico}.{\BBCQ}
\newblock
\APACjournalVolNumPages{Washington DC American Geophysical Union Geophysical
  Monograph Series}{161}{}{123--142}.
\PrintBackRefs{\CurrentBib}

\bibitem [\protect \citeauthoryear {%
Hamilton%
, Lugo-Fern{\'a}ndez%
\BCBL {}\ \BBA {} Sheinbaum%
}{%
Hamilton%
\ \protect \BOthers {.}}{%
{\protect \APACyear {2016}}%
}]{%
hamilton2016loop}
\APACinsertmetastar {%
hamilton2016loop}%
\begin{APACrefauthors}%
Hamilton, P.%
, Lugo-Fern{\'a}ndez, A.%
\BCBL {}\ \BBA {} Sheinbaum, J.%
\end{APACrefauthors}%
\unskip\
\newblock
\APACrefYearMonthDay{2016}{}{}.
\newblock
{\BBOQ}\APACrefatitle {A Loop Current experiment: Field and remote
  measurements} {A loop current experiment: Field and remote
  measurements}.{\BBCQ}
\newblock
\APACjournalVolNumPages{Dynamics of Atmospheres and Oceans}{76}{}{156--173}.
\PrintBackRefs{\CurrentBib}

\bibitem [\protect \citeauthoryear {%
Hu%
\ \protect \BOthers {.}}{%
Hu%
\ \protect \BOthers {.}}{%
{\protect \APACyear {2005}}%
}]{%
hu2005mississippi}
\APACinsertmetastar {%
hu2005mississippi}%
\begin{APACrefauthors}%
Hu, C.%
, Nelson, J\BPBI R.%
, Johns, E.%
, Chen, Z.%
, Weisberg, R\BPBI H.%
\BCBL {}\ \BBA {} M{\"u}ller-Karger, F\BPBI E.%
\end{APACrefauthors}%
\unskip\
\newblock
\APACrefYearMonthDay{2005}{}{}.
\newblock
{\BBOQ}\APACrefatitle {Mississippi River water in the Florida Straits and in
  the Gulf Stream off Georgia in summer 2004} {Mississippi river water in the
  florida straits and in the gulf stream off georgia in summer 2004}.{\BBCQ}
\newblock
\APACjournalVolNumPages{Geophysical Research Letters}{32}{14}{}.
\PrintBackRefs{\CurrentBib}

\bibitem [\protect \citeauthoryear {%
Large%
, McWilliams%
\BCBL {}\ \BBA {} Doney%
}{%
Large%
\ \protect \BOthers {.}}{%
{\protect \APACyear {1994}}%
}]{%
large1994oceanic}
\APACinsertmetastar {%
large1994oceanic}%
\begin{APACrefauthors}%
Large, W\BPBI G.%
, McWilliams, J\BPBI C.%
\BCBL {}\ \BBA {} Doney, S\BPBI C.%
\end{APACrefauthors}%
\unskip\
\newblock
\APACrefYearMonthDay{1994}{}{}.
\newblock
{\BBOQ}\APACrefatitle {Oceanic vertical mixing: A review and a model with a
  nonlocal boundary layer parameterization} {Oceanic vertical mixing: A review
  and a model with a nonlocal boundary layer parameterization}.{\BBCQ}
\newblock
\APACjournalVolNumPages{Reviews of geophysics}{32}{4}{363--403}.
\PrintBackRefs{\CurrentBib}

\bibitem [\protect \citeauthoryear {%
Leben%
}{%
Leben%
}{%
{\protect \APACyear {2005}}%
}]{%
leben2005altimeter}
\APACinsertmetastar {%
leben2005altimeter}%
\begin{APACrefauthors}%
Leben, R\BPBI R.%
\end{APACrefauthors}%
\unskip\
\newblock
\APACrefYearMonthDay{2005}{}{}.
\newblock
{\BBOQ}\APACrefatitle {Altimeter-derived loop current metrics}
  {Altimeter-derived loop current metrics}.{\BBCQ}
\newblock
\APACjournalVolNumPages{Geophysical Monograph-American Geophysical
  Union}{161}{}{181}.
\PrintBackRefs{\CurrentBib}

\bibitem [\protect \citeauthoryear {%
Leben%
\ \BBA {} Born%
}{%
Leben%
\ \BBA {} Born%
}{%
{\protect \APACyear {1993}}%
}]{%
leben1993tracking}
\APACinsertmetastar {%
leben1993tracking}%
\begin{APACrefauthors}%
Leben, R\BPBI R.%
\BCBT {}\ \BBA {} Born, G\BPBI H.%
\end{APACrefauthors}%
\unskip\
\newblock
\APACrefYearMonthDay{1993}{}{}.
\newblock
{\BBOQ}\APACrefatitle {Tracking Loop Current eddies with satellite altimetry}
  {Tracking loop current eddies with satellite altimetry}.{\BBCQ}
\newblock
\APACjournalVolNumPages{Advances in Space Research}{13}{11}{325--333}.
\PrintBackRefs{\CurrentBib}

\bibitem [\protect \citeauthoryear {%
Leben%
, Born%
\BCBL {}\ \BBA {} Engebreth%
}{%
Leben%
\ \protect \BOthers {.}}{%
{\protect \APACyear {2002}}%
}]{%
leben2002operational}
\APACinsertmetastar {%
leben2002operational}%
\begin{APACrefauthors}%
Leben, R\BPBI R.%
, Born, G\BPBI H.%
\BCBL {}\ \BBA {} Engebreth, B\BPBI R.%
\end{APACrefauthors}%
\unskip\
\newblock
\APACrefYearMonthDay{2002}{}{}.
\newblock
{\BBOQ}\APACrefatitle {Operational altimeter data processing for mesoscale
  monitoring} {Operational altimeter data processing for mesoscale
  monitoring}.{\BBCQ}
\newblock
\APACjournalVolNumPages{Marine Geodesy}{25}{1-2}{3--18}.
\PrintBackRefs{\CurrentBib}

\bibitem [\protect \citeauthoryear {%
Lee%
\ \BBA {} Mellor%
}{%
Lee%
\ \BBA {} Mellor%
}{%
{\protect \APACyear {2003}}%
}]{%
lee2003numerical}
\APACinsertmetastar {%
lee2003numerical}%
\begin{APACrefauthors}%
Lee, H\BHBI C.%
\BCBT {}\ \BBA {} Mellor, G\BPBI L.%
\end{APACrefauthors}%
\unskip\
\newblock
\APACrefYearMonthDay{2003}{}{}.
\newblock
{\BBOQ}\APACrefatitle {Numerical simulation of the Gulf Stream System: The Loop
  Current and the deep circulation} {Numerical simulation of the gulf stream
  system: The loop current and the deep circulation}.{\BBCQ}
\newblock
\APACjournalVolNumPages{Journal of Geophysical Research: Oceans}{108}{C2}{}.
\PrintBackRefs{\CurrentBib}

\bibitem [\protect \citeauthoryear {%
Le~H{\'e}naff%
\ \protect \BOthers {.}}{%
Le~H{\'e}naff%
\ \protect \BOthers {.}}{%
{\protect \APACyear {2012}}%
}]{%
le2012surface}
\APACinsertmetastar {%
le2012surface}%
\begin{APACrefauthors}%
Le~H{\'e}naff, M.%
, Kourafalou, V\BPBI H.%
, Paris, C\BPBI B.%
, Helgers, J.%
, Aman, Z\BPBI M.%
, Hogan, P\BPBI J.%
\BCBL {}\ \BBA {} Srinivasan, A.%
\end{APACrefauthors}%
\unskip\
\newblock
\APACrefYearMonthDay{2012}{}{}.
\newblock
{\BBOQ}\APACrefatitle {Surface evolution of the Deepwater Horizon oil spill
  patch: combined effects of circulation and wind-induced drift} {Surface
  evolution of the deepwater horizon oil spill patch: combined effects of
  circulation and wind-induced drift}.{\BBCQ}
\newblock
\APACjournalVolNumPages{Environmental science \&
  technology}{46}{13}{7267--7273}.
\PrintBackRefs{\CurrentBib}

\bibitem [\protect \citeauthoryear {%
Liu%
, Bracco%
\BCBL {}\ \BBA {} Sitar%
}{%
Liu%
, Bracco%
\BCBL {}\ \BBA {} Sitar%
}{%
{\protect \APACyear {2021}}%
}]{%
liu2021submesoscale}
\APACinsertmetastar {%
liu2021submesoscale}%
\begin{APACrefauthors}%
Liu, G.%
, Bracco, A.%
\BCBL {}\ \BBA {} Sitar, A.%
\end{APACrefauthors}%
\unskip\
\newblock
\APACrefYearMonthDay{2021}{}{}.
\newblock
{\BBOQ}\APACrefatitle {Submesoscale Mixing Across the Mixed Layer in the Gulf
  of Mexico} {Submesoscale mixing across the mixed layer in the gulf of
  mexico}.{\BBCQ}
\newblock
\APACjournalVolNumPages{Frontiers in Marine Science}{8}{}{231}.
\PrintBackRefs{\CurrentBib}

\bibitem [\protect \citeauthoryear {%
Liu%
, Falasca%
\BCBL {}\ \BBA {} Bracco%
}{%
Liu%
, Falasca%
\BCBL {}\ \BBA {} Bracco%
}{%
{\protect \APACyear {2021}}%
}]{%
liu2021GRL}
\APACinsertmetastar {%
liu2021GRL}%
\begin{APACrefauthors}%
Liu, G.%
, Falasca, F.%
\BCBL {}\ \BBA {} Bracco, A.%
\end{APACrefauthors}%
\unskip\
\newblock
\APACrefYearMonthDay{2021}{}{}.
\newblock
{\BBOQ}\APACrefatitle {Dynamical Characterization of the Loop Current
  Attractor} {Dynamical characterization of the loop current attractor}.{\BBCQ}
\newblock
\APACjournalVolNumPages{Geophysical Research Letters}{submitted}{}{}.
\PrintBackRefs{\CurrentBib}

\bibitem [\protect \citeauthoryear {%
Luo%
, Bracco%
, Cardona%
\BCBL {}\ \BBA {} McWilliams%
}{%
Luo%
\ \protect \BOthers {.}}{%
{\protect \APACyear {2016}}%
}]{%
luo2016submesoscale}
\APACinsertmetastar {%
luo2016submesoscale}%
\begin{APACrefauthors}%
Luo, H.%
, Bracco, A.%
, Cardona, Y.%
\BCBL {}\ \BBA {} McWilliams, J\BPBI C.%
\end{APACrefauthors}%
\unskip\
\newblock
\APACrefYearMonthDay{2016}{}{}.
\newblock
{\BBOQ}\APACrefatitle {Submesoscale circulation in the northern Gulf of Mexico:
  Surface processes and the impact of the freshwater river input} {Submesoscale
  circulation in the northern gulf of mexico: Surface processes and the impact
  of the freshwater river input}.{\BBCQ}
\newblock
\APACjournalVolNumPages{Ocean Modelling}{101}{}{68--82}.
\PrintBackRefs{\CurrentBib}

\bibitem [\protect \citeauthoryear {%
Mensa%
\ \protect \BOthers {.}}{%
Mensa%
\ \protect \BOthers {.}}{%
{\protect \APACyear {2013}}%
}]{%
mensa2013seasonality}
\APACinsertmetastar {%
mensa2013seasonality}%
\begin{APACrefauthors}%
Mensa, J\BPBI A.%
, Garraffo, Z.%
, Griffa, A.%
, {\"O}zg{\"o}kmen, T\BPBI M.%
, Haza, A.%
\BCBL {}\ \BBA {} Veneziani, M.%
\end{APACrefauthors}%
\unskip\
\newblock
\APACrefYearMonthDay{2013}{}{}.
\newblock
{\BBOQ}\APACrefatitle {Seasonality of the submesoscale dynamics in the Gulf
  Stream region} {Seasonality of the submesoscale dynamics in the gulf stream
  region}.{\BBCQ}
\newblock
\APACjournalVolNumPages{Ocean Dynamics}{63}{8}{923--941}.
\PrintBackRefs{\CurrentBib}

\bibitem [\protect \citeauthoryear {%
Mooers%
, Zaron%
\BCBL {}\ \BBA {} Howard%
}{%
Mooers%
\ \protect \BOthers {.}}{%
{\protect \APACyear {2012}}%
}]{%
mooers2012final}
\APACinsertmetastar {%
mooers2012final}%
\begin{APACrefauthors}%
Mooers, C.%
, Zaron, E.%
\BCBL {}\ \BBA {} Howard, M.%
\end{APACrefauthors}%
\unskip\
\newblock
\APACrefYearMonthDay{2012}{}{}.
\newblock
{\BBOQ}\APACrefatitle {Final report for phase i: Gulf of mexico 3-d operational
  ocean forecast system pilot prediction project (gomex-ppp)} {Final report for
  phase i: Gulf of mexico 3-d operational ocean forecast system pilot
  prediction project (gomex-ppp)}.{\BBCQ}
\newblock
\APACjournalVolNumPages{Final Rep. to Research Partnership to Secure Energy for
  America}{149}{}{}.
\PrintBackRefs{\CurrentBib}

\bibitem [\protect \citeauthoryear {%
Moore%
}{%
Moore%
}{%
{\protect \APACyear {1999}}%
}]{%
moore1999dynamics}
\APACinsertmetastar {%
moore1999dynamics}%
\begin{APACrefauthors}%
Moore, A\BPBI M.%
\end{APACrefauthors}%
\unskip\
\newblock
\APACrefYearMonthDay{1999}{}{}.
\newblock
{\BBOQ}\APACrefatitle {The dynamics of error growth and predictability in a
  model of the Gulf Stream. Part II: Ensemble prediction} {The dynamics of
  error growth and predictability in a model of the gulf stream. part ii:
  Ensemble prediction}.{\BBCQ}
\newblock
\APACjournalVolNumPages{Journal of physical oceanography}{29}{4}{762--778}.
\PrintBackRefs{\CurrentBib}

\bibitem [\protect \citeauthoryear {%
Murphy%
}{%
Murphy%
}{%
{\protect \APACyear {1988}}%
}]{%
murphy1988impact}
\APACinsertmetastar {%
murphy1988impact}%
\begin{APACrefauthors}%
Murphy, J\BPBI M.%
\end{APACrefauthors}%
\unskip\
\newblock
\APACrefYearMonthDay{1988}{}{}.
\newblock
{\BBOQ}\APACrefatitle {The impact of ensemble forecasts on predictability} {The
  impact of ensemble forecasts on predictability}.{\BBCQ}
\newblock
\APACjournalVolNumPages{Quarterly Journal of the Royal Meteorological
  Society}{114}{480}{463--493}.
\PrintBackRefs{\CurrentBib}

\bibitem [\protect \citeauthoryear {%
{National Academies of Sciences, Engineering and Medicine}%
}{%
{National Academies of Sciences, Engineering and Medicine}%
}{%
{\protect \APACyear {2018}}%
}]{%
NAS2018report}
\APACinsertmetastar {%
NAS2018report}%
\begin{APACrefauthors}%
{National Academies of Sciences, Engineering and Medicine}.%
\end{APACrefauthors}%
\unskip\
\newblock
\APACrefYear{2018}.
\newblock
\APACrefbtitle {Understanding and Predicting the Gulf of Mexico Loop Current:
  Critical Gaps and Recommendations} {Understanding and predicting the gulf of
  mexico loop current: Critical gaps and recommendations}.
\newblock
\APACaddressPublisher{Washington, DC}{The National Academies Press}.
\PrintBackRefs{\CurrentBib}

\bibitem [\protect \citeauthoryear {%
L.~Oey%
}{%
L.~Oey%
}{%
{\protect \APACyear {2005}}%
}]{%
oey2005wetting}
\APACinsertmetastar {%
oey2005wetting}%
\begin{APACrefauthors}%
Oey, L.%
\end{APACrefauthors}%
\unskip\
\newblock
\APACrefYearMonthDay{2005}{}{}.
\newblock
{\BBOQ}\APACrefatitle {A wetting and drying scheme for POM} {A wetting and
  drying scheme for pom}.{\BBCQ}
\newblock
\APACjournalVolNumPages{Ocean Modelling}{9}{2}{133--150}.
\PrintBackRefs{\CurrentBib}

\bibitem [\protect \citeauthoryear {%
L.~Oey%
, Ezer%
\BCBL {}\ \BBA {} Lee%
}{%
L.~Oey%
\ \protect \BOthers {.}}{%
{\protect \APACyear {2005}}%
}]{%
oey2005loop}
\APACinsertmetastar {%
oey2005loop}%
\begin{APACrefauthors}%
Oey, L.%
, Ezer, T.%
\BCBL {}\ \BBA {} Lee, H.%
\end{APACrefauthors}%
\unskip\
\newblock
\APACrefYearMonthDay{2005}{}{}.
\newblock
{\BBOQ}\APACrefatitle {Loop Current, rings and related circulation in the Gulf
  of Mexico: A review of numerical models and future challenges} {Loop current,
  rings and related circulation in the gulf of mexico: A review of numerical
  models and future challenges}.{\BBCQ}
\newblock
\APACjournalVolNumPages{Geophysical Monograph-American Geophysical
  Union}{161}{}{31}.
\PrintBackRefs{\CurrentBib}

\bibitem [\protect \citeauthoryear {%
L\BHBI Y.~Oey%
}{%
L\BHBI Y.~Oey%
}{%
{\protect \APACyear {1996}}%
}]{%
oey1996simulation}
\APACinsertmetastar {%
oey1996simulation}%
\begin{APACrefauthors}%
Oey, L\BHBI Y.%
\end{APACrefauthors}%
\unskip\
\newblock
\APACrefYearMonthDay{1996}{}{}.
\newblock
{\BBOQ}\APACrefatitle {Simulation of mesoscale variability in the Gulf of
  Mexico: Sensitivity studies, comparison with observations, and trapped wave
  propagation} {Simulation of mesoscale variability in the gulf of mexico:
  Sensitivity studies, comparison with observations, and trapped wave
  propagation}.{\BBCQ}
\newblock
\APACjournalVolNumPages{Journal of physical Oceanography}{26}{2}{145--174}.
\PrintBackRefs{\CurrentBib}

\bibitem [\protect \citeauthoryear {%
L\BHBI Y.~Oey%
}{%
L\BHBI Y.~Oey%
}{%
{\protect \APACyear {2004}}%
}]{%
oey2004vorticity}
\APACinsertmetastar {%
oey2004vorticity}%
\begin{APACrefauthors}%
Oey, L\BHBI Y.%
\end{APACrefauthors}%
\unskip\
\newblock
\APACrefYearMonthDay{2004}{}{}.
\newblock
{\BBOQ}\APACrefatitle {Vorticity flux through the Yucatan Channel and Loop
  Current variability in the Gulf of Mexico} {Vorticity flux through the
  yucatan channel and loop current variability in the gulf of mexico}.{\BBCQ}
\newblock
\APACjournalVolNumPages{Journal of Geophysical Research: Oceans}{109}{C10}{}.
\PrintBackRefs{\CurrentBib}

\bibitem [\protect \citeauthoryear {%
Poje%
\ \protect \BOthers {.}}{%
Poje%
\ \protect \BOthers {.}}{%
{\protect \APACyear {2014}}%
}]{%
poje2014submesoscale}
\APACinsertmetastar {%
poje2014submesoscale}%
\begin{APACrefauthors}%
Poje, A\BPBI C.%
, {\"O}zg{\"o}kmen, T\BPBI M.%
, Lipphardt, B\BPBI L.%
, Haus, B\BPBI K.%
, Ryan, E\BPBI H.%
, Haza, A\BPBI C.%
\BDBL {}others%
\end{APACrefauthors}%
\unskip\
\newblock
\APACrefYearMonthDay{2014}{}{}.
\newblock
{\BBOQ}\APACrefatitle {Submesoscale dispersion in the vicinity of the Deepwater
  Horizon spill} {Submesoscale dispersion in the vicinity of the deepwater
  horizon spill}.{\BBCQ}
\newblock
\APACjournalVolNumPages{Proceedings of the National Academy of
  Sciences}{111}{35}{12693--12698}.
\PrintBackRefs{\CurrentBib}

\bibitem [\protect \citeauthoryear {%
Sandery%
\ \BBA {} Sakov%
}{%
Sandery%
\ \BBA {} Sakov%
}{%
{\protect \APACyear {2017}}%
}]{%
sandery2017ocean}
\APACinsertmetastar {%
sandery2017ocean}%
\begin{APACrefauthors}%
Sandery, P\BPBI A.%
\BCBT {}\ \BBA {} Sakov, P.%
\end{APACrefauthors}%
\unskip\
\newblock
\APACrefYearMonthDay{2017}{}{}.
\newblock
{\BBOQ}\APACrefatitle {Ocean forecasting of mesoscale features can deteriorate
  by increasing model resolution towards the submesoscale} {Ocean forecasting
  of mesoscale features can deteriorate by increasing model resolution towards
  the submesoscale}.{\BBCQ}
\newblock
\APACjournalVolNumPages{Nature communications}{8}{1}{1--8}.
\PrintBackRefs{\CurrentBib}

\bibitem [\protect \citeauthoryear {%
Schiller%
\ \BBA {} Kourafalou%
}{%
Schiller%
\ \BBA {} Kourafalou%
}{%
{\protect \APACyear {2014}}%
}]{%
schiller2014loop}
\APACinsertmetastar {%
schiller2014loop}%
\begin{APACrefauthors}%
Schiller, R.%
\BCBT {}\ \BBA {} Kourafalou, V.%
\end{APACrefauthors}%
\unskip\
\newblock
\APACrefYearMonthDay{2014}{}{}.
\newblock
{\BBOQ}\APACrefatitle {Loop Current impact on the transport of Mississippi
  River waters} {Loop current impact on the transport of mississippi river
  waters}.{\BBCQ}
\newblock
\APACjournalVolNumPages{Journal of Coastal Research}{30}{6}{1287--1306}.
\PrintBackRefs{\CurrentBib}

\bibitem [\protect \citeauthoryear {%
Schiller%
, Kourafalou%
, Hogan%
\BCBL {}\ \BBA {} Walker%
}{%
Schiller%
\ \protect \BOthers {.}}{%
{\protect \APACyear {2011}}%
}]{%
schiller2011dynamics}
\APACinsertmetastar {%
schiller2011dynamics}%
\begin{APACrefauthors}%
Schiller, R.%
, Kourafalou, V\BPBI H.%
, Hogan, P.%
\BCBL {}\ \BBA {} Walker, N.%
\end{APACrefauthors}%
\unskip\
\newblock
\APACrefYearMonthDay{2011}{}{}.
\newblock
{\BBOQ}\APACrefatitle {The dynamics of the Mississippi River plume: Impact of
  topography, wind and offshore forcing on the fate of plume waters} {The
  dynamics of the mississippi river plume: Impact of topography, wind and
  offshore forcing on the fate of plume waters}.{\BBCQ}
\newblock
\APACjournalVolNumPages{Journal of Geophysical Research: Oceans}{116}{C6}{}.
\PrintBackRefs{\CurrentBib}

\bibitem [\protect \citeauthoryear {%
Shchepetkin%
\ \BBA {} McWilliams%
}{%
Shchepetkin%
\ \BBA {} McWilliams%
}{%
{\protect \APACyear {2005}}%
}]{%
shchepetkin2005regional}
\APACinsertmetastar {%
shchepetkin2005regional}%
\begin{APACrefauthors}%
Shchepetkin, A\BPBI F.%
\BCBT {}\ \BBA {} McWilliams, J\BPBI C.%
\end{APACrefauthors}%
\unskip\
\newblock
\APACrefYearMonthDay{2005}{}{}.
\newblock
{\BBOQ}\APACrefatitle {The regional oceanic modeling system (ROMS): a
  split-explicit, free-surface, topography-following-coordinate oceanic model}
  {The regional oceanic modeling system (roms): a split-explicit, free-surface,
  topography-following-coordinate oceanic model}.{\BBCQ}
\newblock
\APACjournalVolNumPages{Ocean modelling}{9}{4}{347--404}.
\PrintBackRefs{\CurrentBib}

\bibitem [\protect \citeauthoryear {%
Sheremet%
, Khan%
\BCBL {}\ \BBA {} Kuehl%
}{%
Sheremet%
\ \protect \BOthers {.}}{%
{\protect \APACyear {2021}}%
}]{%
Sheremet2021JPO}
\APACinsertmetastar {%
Sheremet2021JPO}%
\begin{APACrefauthors}%
Sheremet, V\BPBI A.%
, Khan, A\BPBI A.%
\BCBL {}\ \BBA {} Kuehl, J\BPBI J.%
\end{APACrefauthors}%
\unskip\
\newblock
\APACrefYearMonthDay{2021}{}{}.
\newblock
{\BBOQ}\APACrefatitle {Multiple equilibrium states of the Loop Current in the
  Gulf of Mexico} {Multiple equilibrium states of the loop current in the gulf
  of mexico}.{\BBCQ}
\newblock
\APACjournalVolNumPages{Journal of Physical Oceanography}{submitted}{}{}.
\PrintBackRefs{\CurrentBib}

\bibitem [\protect \citeauthoryear {%
Stewart%
\ \protect \BOthers {.}}{%
Stewart%
\ \protect \BOthers {.}}{%
{\protect \APACyear {2017}}%
}]{%
stewart2017vertical}
\APACinsertmetastar {%
stewart2017vertical}%
\begin{APACrefauthors}%
Stewart, K.%
, Hogg, A\BPBI M.%
, Griffies, S.%
, Heerdegen, A.%
, Ward, M.%
, Spence, P.%
\BCBL {}\ \BBA {} England, M.%
\end{APACrefauthors}%
\unskip\
\newblock
\APACrefYearMonthDay{2017}{}{}.
\newblock
{\BBOQ}\APACrefatitle {Vertical resolution of baroclinic modes in global ocean
  models} {Vertical resolution of baroclinic modes in global ocean
  models}.{\BBCQ}
\newblock
\APACjournalVolNumPages{Ocean Modelling}{113}{}{50--65}.
\PrintBackRefs{\CurrentBib}

\bibitem [\protect \citeauthoryear {%
Sturges%
\ \BBA {} Leben%
}{%
Sturges%
\ \BBA {} Leben%
}{%
{\protect \APACyear {2000}}%
}]{%
sturges2000frequency}
\APACinsertmetastar {%
sturges2000frequency}%
\begin{APACrefauthors}%
Sturges, W.%
\BCBT {}\ \BBA {} Leben, R.%
\end{APACrefauthors}%
\unskip\
\newblock
\APACrefYearMonthDay{2000}{}{}.
\newblock
{\BBOQ}\APACrefatitle {Frequency of ring separations from the Loop Current in
  the Gulf of Mexico: A revised estimate} {Frequency of ring separations from
  the loop current in the gulf of mexico: A revised estimate}.{\BBCQ}
\newblock
\APACjournalVolNumPages{Journal of Physical Oceanography}{30}{7}{1814--1819}.
\PrintBackRefs{\CurrentBib}

\bibitem [\protect \citeauthoryear {%
Sun%
\ \protect \BOthers {.}}{%
Sun%
\ \protect \BOthers {.}}{%
{\protect \APACyear {2020}}%
}]{%
sun2020diurnal}
\APACinsertmetastar {%
sun2020diurnal}%
\begin{APACrefauthors}%
Sun, D.%
, Bracco, A.%
, Barkan, R.%
, Berta, M.%
, Dauhajre, D.%
, Molemaker, M\BPBI J.%
\BDBL {}McWilliams, J\BPBI C.%
\end{APACrefauthors}%
\unskip\
\newblock
\APACrefYearMonthDay{2020}{}{}.
\newblock
{\BBOQ}\APACrefatitle {Diurnal cycling of submesoscale dynamics: Lagrangian
  implications in drifter observations and model simulations of the northern
  gulf of mexico} {Diurnal cycling of submesoscale dynamics: Lagrangian
  implications in drifter observations and model simulations of the northern
  gulf of mexico}.{\BBCQ}
\newblock
\APACjournalVolNumPages{Journal of Physical Oceanography}{50}{6}{1605--1623}.
\PrintBackRefs{\CurrentBib}

\bibitem [\protect \citeauthoryear {%
Thoppil%
\ \protect \BOthers {.}}{%
Thoppil%
\ \protect \BOthers {.}}{%
{\protect \APACyear {2021}}%
}]{%
thoppil2021ensemble}
\APACinsertmetastar {%
thoppil2021ensemble}%
\begin{APACrefauthors}%
Thoppil, P\BPBI G.%
, Frolov, S.%
, Rowley, C\BPBI D.%
, Reynolds, C\BPBI A.%
, Jacobs, G\BPBI A.%
, Metzger, E\BPBI J.%
\BDBL {}others%
\end{APACrefauthors}%
\unskip\
\newblock
\APACrefYearMonthDay{2021}{}{}.
\newblock
{\BBOQ}\APACrefatitle {Ensemble forecasting greatly expands the prediction
  horizon for ocean mesoscale variability} {Ensemble forecasting greatly
  expands the prediction horizon for ocean mesoscale variability}.{\BBCQ}
\newblock
\APACjournalVolNumPages{Communications Earth \& Environment}{2}{1}{1--9}.
\PrintBackRefs{\CurrentBib}

\bibitem [\protect \citeauthoryear {%
Vukovich%
}{%
Vukovich%
}{%
{\protect \APACyear {2007}}%
}]{%
vukovich2007climatology}
\APACinsertmetastar {%
vukovich2007climatology}%
\begin{APACrefauthors}%
Vukovich, F\BPBI M.%
\end{APACrefauthors}%
\unskip\
\newblock
\APACrefYearMonthDay{2007}{}{}.
\newblock
{\BBOQ}\APACrefatitle {Climatology of ocean features in the Gulf of Mexico
  using satellite remote sensing data} {Climatology of ocean features in the
  gulf of mexico using satellite remote sensing data}.{\BBCQ}
\newblock
\APACjournalVolNumPages{Journal of Physical Oceanography}{37}{3}{689--707}.
\PrintBackRefs{\CurrentBib}

\bibitem [\protect \citeauthoryear {%
Yin%
\ \BBA {} Oey%
}{%
Yin%
\ \BBA {} Oey%
}{%
{\protect \APACyear {2007}}%
}]{%
yin2007bred}
\APACinsertmetastar {%
yin2007bred}%
\begin{APACrefauthors}%
Yin, X\BHBI Q.%
\BCBT {}\ \BBA {} Oey, L\BHBI Y.%
\end{APACrefauthors}%
\unskip\
\newblock
\APACrefYearMonthDay{2007}{}{}.
\newblock
{\BBOQ}\APACrefatitle {Bred-ensemble ocean forecast of Loop Current and rings}
  {Bred-ensemble ocean forecast of loop current and rings}.{\BBCQ}
\newblock
\APACjournalVolNumPages{Ocean Modelling}{17}{4}{300--326}.
\PrintBackRefs{\CurrentBib}

\bibitem [\protect \citeauthoryear {%
Zhong%
\ \BBA {} Bracco%
}{%
Zhong%
\ \BBA {} Bracco%
}{%
{\protect \APACyear {2013}}%
}]{%
zhong2013submesoscale}
\APACinsertmetastar {%
zhong2013submesoscale}%
\begin{APACrefauthors}%
Zhong, Y.%
\BCBT {}\ \BBA {} Bracco, A.%
\end{APACrefauthors}%
\unskip\
\newblock
\APACrefYearMonthDay{2013}{}{}.
\newblock
{\BBOQ}\APACrefatitle {Submesoscale impacts on horizontal and vertical
  transport in the Gulf of Mexico} {Submesoscale impacts on horizontal and
  vertical transport in the gulf of mexico}.{\BBCQ}
\newblock
\APACjournalVolNumPages{Journal of Geophysical Research:
  Oceans}{118}{10}{5651--5668}.
\PrintBackRefs{\CurrentBib}

\bibitem [\protect \citeauthoryear {%
Zhong%
, Bracco%
\BCBL {}\ \BBA {} Villareal%
}{%
Zhong%
\ \protect \BOthers {.}}{%
{\protect \APACyear {2012}}%
}]{%
zhong2012pattern}
\APACinsertmetastar {%
zhong2012pattern}%
\begin{APACrefauthors}%
Zhong, Y.%
, Bracco, A.%
\BCBL {}\ \BBA {} Villareal, T\BPBI A.%
\end{APACrefauthors}%
\unskip\
\newblock
\APACrefYearMonthDay{2012}{}{}.
\newblock
{\BBOQ}\APACrefatitle {Pattern formation at the ocean surface: Sargassum
  distribution and the role of the eddy field} {Pattern formation at the ocean
  surface: Sargassum distribution and the role of the eddy field}.{\BBCQ}
\newblock
\APACjournalVolNumPages{Limnology and Oceanography: Fluids and
  Environments}{2}{1}{12--27}.
\PrintBackRefs{\CurrentBib}

\end{thebibliography}

%
%
%
%
%

\end{document}